\documentclass[useAMS,usenatbib]{mn2e}
\usepackage{bibentry}
\usepackage{amsmath}
\usepackage{mathtools}
\usepackage{float}
\usepackage{amssymb}
\usepackage{epsfig}
\usepackage{color}
\usepackage{ulem}
\newcommand{\HI}{\textsc{Hi}\,}
\newcommand{\HII}{\textsc{Hii}}

\newcommand{\aj}{AJ}
\newcommand{\apj}{ApJ}
\newcommand{\apjl}{ApJL}
\newcommand{\apjs}{ApJS}
\newcommand{\mnras}{MNRAS}
\newcommand{\aap}{A\&A}
\newcommand{\araa}{ARA\&A}

\newcommand{\gsim}{\lower.7ex\hbox{$\;\stackrel{\textstyle>}{\sim}\;$}}
\newcommand{\lsim}{\lower.7ex\hbox{$\;\stackrel{\textstyle<}{\sim}\;$}}
\def\ltsima{$\; \buildrel < \over \sim \;$}
\def\simlt{\lower.5ex\hbox{\ltsima}}
\def\gtsima{$\; \buildrel > \over \sim \;$}
\def\simgt{\lower.5ex\hbox{\gtsima}}
\def\la{$\langle$}

\title[Simulating H$_2$ in high-$z$ galaxies]{Simulating the H$_2$ content of high-redshift galaxies}

\author[Tomassetti et al.]{Matteo Tomassetti
\thanks{E-mail:mtomas@astro.uni-bonn.de \newline
  Member of the International Max Planck Research School (IMPRS) for
  Astronomy and Astrophysics at the Universities of Bonn and Cologne},
  Cristiano Porciani, Emilio Romano-D\'{\i}az, Aaron D. Ludlow\\
  Argelander Institut f\"{u}r Astronomie, Auf Dem H\"{u}gel 71, D-53121 Bonn, Germany\\
  }

\begin{document}

\date{Accepted 2014 October 28. Received 2014 October 20; in original form 2014 March 26}

\pagerange{\pageref{firstpage}--\pageref{lastpage}} \pubyear{2014}

\maketitle

\label{firstpage}

\begin{abstract}
We introduce a sub-grid model for the non-equilibrium abundance of 
molecular hydrogen in cosmological simulations of galaxy formation. 
We improve upon previous work by accounting for the unresolved structure of 
molecular clouds in a phenomenological way which combines both observational 
and numerical results on the properties of the turbulent interstellar medium.
We apply the model to a cosmological simulation of the formation of a 
Milky Way-sized galaxy at $z=2$, and compare the results to those obtained 
using other popular prescriptions that compute the equilibrium abundance of 
H$_2$. In these runs we introduce an explicit link between star formation 
and the local H$_2$ abundance, and perform an additional simulation in which 
star formation is linked directly to the density of cold gas.
In better agreement with observations, we find that the simulated galaxy produces 
less stars and harbours a larger gas reservoir when star formation is regulated by 
molecular hydrogen. In this case, the galaxy is composed of a younger stellar 
population as early star formation is inhibited in small, metal poor dark-matter 
haloes which cannot efficiently produce H$_2$. The number of luminous satellites 
orbiting within the virial radius of the galaxy at $z=2$ is reduced by 10-30 
per cent in models with H$_2$-regulated star formation.
\end{abstract}


\section{Introduction}

The process of galaxy formation involves the interplay of many non-linear phenomena that span a 
wide range of length and time-scales. A galaxy like our Milky Way, for example, forms from a region 
that initially extends to roughly one comoving Mpc, yet its angular momentum is determined 
by the mass distribution within tens of comoving Mpc. Star
formation (SF), on the other hand, takes place in the densest cores of giant molecular clouds 
(GMCs), on scales of the order of 0.1 pc. 

The challenge in simulations of galaxy formation is to capture this vast dynamic
range, while simultaneously accounting for the different physical processes that intervene on
relevant scales. This is 
usually achieved with ad hoc sub-grid models that attempt to emulate the most
important small-scale phenomena. In particular, one of the biggest uncertainties in simulations of
galaxy formation is the means by which gas is converted into stars 
\citep[see][for a recent review]{Dobbs+13}. 

The standard approach to this problem, motivated by observations, is to adopt a Schmidt-like 
law \citep{Schmidt+59}, often coupled to conditions on the local gas properties.
However, there are several issues with this method. First, its parameters 
are poorly constrained and are usually fine-tuned to match the observed Kennicutt-Schmidt 
(KS) relation \citep{Kennicutt+89, Kennicutt+98}. Secondly, there is a growing body of 
evidence that the local star formation rate (SFR) correlates more tightly with the observed 
density of molecular hydrogen than with that of the total gas density 
\citep[e.g.][]{Kennicutt+07,Bigiel+08,Leroy+08}, though there is yet no consensus as to
whether this reflects a causal relation.
In particular, numerical simulations of isolated molecular clouds suggest
that the presence of molecules does not boost the ability of the gas to cool
and form stars \citep{Glover+12b}. 
The tight spatial correlation between H$_2$ and young stars may then be due 
to the fact that they are both formed in high density regions where gas is 
effectively shielded from the interstellar radiation field.

Despite the ongoing debate,
there are strong motivations for including a treatment of molecular hydrogen
in cosmological simulations of galaxy formation. Observations of 
H$_2$ proxies (such as CO luminosity), for example, have progressed tremendously over the
past decade \citep[see][for a recent review]{Carilli+13}, underlining the need
for robust theoretical templates to aid in the design of observational campaigns
and the interpretation of their results. Furthermore, numerical simulations constitute a unique 
tool to test the impact of H$_2$-regulated SF on the global structure of 
galaxies, provided their H$_2$ content can be reliably determined.

Tracking H$_2$, however, requires solving a challenging network of rate equations which 
are coupled to a radiative-transfer computation for H$_2$-dissociating photons. Given 
that the spatial resolution of current simulations is comparable in size to GMCs, these 
calculations must be done at the sub-grid level and include a description of gas structure 
on the unresolved scales (e.g. a clumping factor for the gas density).

Recently, several authors have incorporated simple algorithms to track   
molecular hydrogen in hydrodynamical simulations of galaxy formation.
For instance, \citet{Pelupessy+06} monitored 
the H$_2$ distribution in dwarf-sized galaxies within a fixed dark matter (DM) potential 
and showed that the resulting molecular mass depends strongly on the metallicity 
of the interstellar medium (ISM). Similar conclusions were drawn by 
\citet[][see also Feldmann et al. (2011)]{Gnedin+09} \nocite{Feldmann+2011}, who
followed the evolution of the H$_2$ content for 100 Myr in a cosmologically simulated galaxy at $z=4$. 
These authors showed that it is only 
possible to form fully shielded molecular clouds when the gas metallicity is 
high (i.e. $Z\sim 10^{-2}-10^{-1}$ Z$_\odot$), and argued that 
H$_2$-regulated SF can act as an effective feedback mechanism,
delaying SF in the low-metallicity progenitors of a galaxy.

The implications of these results for galaxy formation in low-mass haloes
were studied further by \citet{Kuhlen+12,Kuhlen+13}, who suggested  
the possible existence of a large population of low-mass, gas-rich galaxies 
that never reached the critical column density required for the H$_2$/\HI\,
transition and are thus devoid of stars. Their work, however, was based on an
analytic model for H$_2$ formation that assumes chemical equilibrium between its 
formation and destruction rates \citep{Krumholz+08,Krumholz+09b,McKee+10}. 
None the less, \citet{Krumholz+11} showed that this model agrees well with a
time-dependent solution to the chemical network provided the local metallicity 
of the gas is above $10^{-2}$ Z$_\odot$, lending support to these conclusions.

\citet{Christensen+12} modelled the non-equilibrium abundance of H$_2$ in a dwarf 
galaxy that was simulated down to redshift $z=0$, connecting SF explicitly 
to the local H$_2$ content of the gas. These authors showed that, compared to simulations 
rooted on the Schmidt law, molecule-based SF produces a galaxy which is
more gas rich, has bluer stellar populations and a clumpier ISM. On the other hand, 
strong stellar feedback, when included, tends to mitigate these differences by regulating 
the formation and destruction rates of GMCs \citep{Hopkins+12}. 

In this paper, we introduce a new time-dependent sub-grid model for tracking the 
non-equilibrium abundance of H$_2$ in cosmological simulations of galaxy assembly.
Our approach builds upon the work of \citet{Gnedin+09} and \citet{Christensen+12} 
by including additional information on the unresolved distribution of gas temperatures 
and densities. In particular, our model: (i) explicitly accounts for the distribution 
of sub-grid densities, as determined by observations and numerical simulations of 
turbulent GMCs; (ii) invokes a gas temperature-density relation that was determined 
from detailed numerical studies of the ISM \citep{Glover+07a}; and (iii) consistently 
takes into account that denser, unresolved clumps have larger optical depths.

As an application, we employ the model in a high-resolution simulation that follows the 
formation of a Milky Way-sized galaxy down to $z=2$. In order to explore the interplay 
between SF, H$_2$ abundance and galactic structure, we re-simulate the same 
volume using different algorithms for computing the density of molecular hydrogen and the 
local SFR.

This paper is organized as follows. In Section \ref{sec:mol_hydrogen_formation}, we introduce 
our model for tracking the non-equilibrium  H$_2$ abundance and compare it with other commonly 
adopted prescriptions that have been discussed in the literature. In Sections \ref{sec:code} 
and \ref{sec:analysis}, we describe our suite of simulations and present our main results. 
Finally, we summarize our main conclusions and then critically discuss some of our assumptions 
in Section \ref{sec:discussion}.

\section{Molecular hydrogen}\label{sec:mol_hydrogen_formation}

The abundance of molecular hydrogen in the metal-rich ISM is mainly regulated by the 
competition between its formation due to the catalytic action of dust grains and its 
photo-dissociation via the two-step Solomon process. In the ground state, H$_2$ absorbs 
electromagnetic radiation in two densely packed series of lines (the Lyman band -- 
characterized by photon energies $E>11.2$ eV or wavelengths $\lambda<1108$ \AA\, -- 
and the Werner band -- $E>12.3$ eV, $\lambda<1008$ \AA). Radiative decay from the 
excited states leads to dissociation in approximately 15 per cent of the cases.
Direct photo-dissociation would require photons with energy $E>14.7$ eV,
but these are principally absorbed by hydrogen atoms as they lie above the hydrogen 
photo-ionization threshold (13.6 eV, 912 \AA). As a result, only photons between 
$912\ \text{\AA}<\lambda<1108\ \text{\AA}$ can photo-dissociate molecular hydrogen.

Lyman-Werner (LW) photons are copiously emitted by OB stars, but intervening H$_2$ 
and dust effectively shield the densest regions of the ISM. This results in an H$_2$ 
abundance that increases rapidly towards regions in which the medium becomes optically 
thick to LW radiation.

An exact treatment of these effects is challenging: it requires three-dimensional radiative 
transfer calculations capable of resolving length and time-scales orders of magnitude shorter 
than those associated with galaxy evolution. Nevertheless, it is possible to follow the formation 
of molecular complexes in a phenomenological way, using approximate sub-grid treatments of the 
most crucial physical processes involved.

In this work, we use three different mathematical models that attempt to approximate these 
effects. The simplest (KMT-EQ) is fully analytical; it returns the equilibrium H$_2$ fraction 
in terms of quantities that can be determined locally in a simulation 
\citep{Krumholz+08,Krumholz+09b,McKee+10}. The model is based on a spherical molecular complex 
immersed in an isotropic bath of LW photons. Assuming that the ISM is in a two-phase equilibrium 
between a cold and a warm neutral medium, it allows both the intensity of the radiation field 
and the resulting H$_2$ fraction to be expressed in terms of the local gas column density and 
metallicity \citep{Krumholz+11}.

The second model (KMT-UV) employs the same analytical relation between the equilibrium 
H$_2$ fraction and the radiation density in the LW band. However, in this case, the latter 
is calculated numerically within the simulations themselves. This is achieved by propagating 
photons from stellar particles assuming that the ISM is optically thin within a characteristic
length scale and optically thick beyond that (see Appendix~\ref{sec:uv} for further details).

Note, however, that the formation of molecular hydrogen on dust grains is a very
inefficient process. In the metal poor ISM, for example, the H$_2$ formation
time-scale approaches a Hubble time. This calls into question the appropriateness of the 
equilibrium-based models, which \textit{instantaneously} populate each simulation volume element 
with a given fraction of H$_2$. This has motivated several authors \citep{Pelupessy+06,Gnedin+09,Christensen+12} 
to develop more sophisticated algorithms capable of tracking the non-equilibrium H$_2$ fraction
in high-resolution simulations. Inspired by these efforts, we have developed a novel dynamical model 
(DYN) for calculating the abundance of molecular hydrogen. This model is described in detail in the 
following subsection.

\subsection{Mathematical formulation of the model}\label{sec:DYN}
The evolution of the H$_2$ number density is described by the following
system of equations:
\begin{align}
&\label{eq_form_H2}\dfrac{{\rm d} n_{\rm H_2}}{{\rm d}t}  = \mathcal{R}_f(T) \, n_{\rm HI} \, n_{\rm H}-G \, \kappa\, \Phi\, e^{-\tau}\,   n_{\rm H_2}  \\
      &\qquad\qquad\qquad\qquad\qquad - \gamma_{\rm HI} \, n_{\rm HI}\, n_{\rm H_2}-\gamma_{\rm H_2} \, n_{\rm H_2}^2  \nonumber\,, 
      \\
&\label{eq_form_HI}\dfrac{{{\rm d} n_\HI}}{{\rm d}t} = n_e \, n_\HII \, k_1(T)   \\
      &\qquad\qquad-n_{\rm HI} \, \left[ k_2(T) \, n_e+\Gamma(z)\right]-2 \, \dfrac{{\rm d}n_{\rm H_2}}{{\rm d}t} \nonumber\, ,\\
&\label{eq_closure_H} n_\HI+n_\HII+2n_{\rm H_2} = n_{\rm H}\,\,\,.
\end{align}
Here $n_{\rm HI}$, $n_\HII$ and $n_{\rm H_2}$ are, respectively,
the number densities of neutral, ionized and molecular hydrogen; $n_e$
is the electron number density; $\gamma_i$ is the ${\rm H_2}$ collisional
destruction rate due to interaction with species $i$; 
$G$ is the unshielded interstellar UV-radiation flux in Habing units 
(see Appendix \ref{sec:uv}); 
$\kappa$ is the
H$_2$ photo-dissociation rate for $G=1$; $\Gamma(z)$ is the photo-ionization 
rate\footnote{In terms of the UV-background intensity, $J_\nu$, the ionization 
  cross-section $\sigma$ and the corresponding optical depth $\tau'$, 
  $\Gamma(z) = \int_{\nu_0}^{+\infty}4\pi J(\nu,z)\,\sigma(\nu)\,e^{-\tau '(\nu)}/
  (h_{\rm p}\,\nu)\,{\rm d}\nu$, where $\nu_0={\rm 13.6\ eV}/h_{\rm p}$ and $h_{\rm p}$ 
  denotes the Planck constant.}
of ${\rm HI}$; $k_1$ and
$k_2$ are the hydrogen recombination and collisional destruction
rates; $\tau =\sigma_{d} N_{\rm H}$ is the optical depth of dust in
the LW bands (conventionally evaluated at 1000 \AA), 
where $N_{\rm H}$ is the total hydrogen column density. 
This assumes that the dust abundance scales linearly with the gas
metallicity and the dust-to-gas mass ratio is equal to the value
measured in the Milky Way (MW), i.e. 10$^{-2}$ $Z/$Z$_\odot$. 
The parametrization of the H$_2$ self-shielding function, 
$\Phi$ (approximated for the plane parallel case) and the H$_2$ formation 
rate on dust grains, $\mathcal{R}_f$, are given in Table \ref{table:pararameters},
and adopt a dust temperature $T_d=10$ K.

\begin{table*}
\centering
\begin{tabular}{l|ccc}
\hline
Parameter & Symbol & Expression & Reference \\
\hline
Clumping factor & $C_\rho$& $\langle\ n_{\rm H}^2\rangle/\langle\ n_{\rm H}\rangle^2=10$ & 1,2 \\
H$_2$ formation rate on dust grains & $\mathcal{R}_f(T)$ & $3.025\times 10^{-17}S_{\rm H}(T)({T}/{\rm 100\,K})^{0.5}\,(Z /\rm Z_\odot)$ & 5,6 \\
Sticking probability for H atoms& $S_{\rm H}(T)$& $  \left[1+0.4\left(\frac{T+T_d}{\rm 100\,K}\right)^{1/2}+0.2\left(\frac{T}{\rm 100\,K}\right)+0.08\left(\frac{T}{\rm 100\,K}\right)^2\right]^{-1} $, $T_d=10$ K & 3,4\\
H$_2$ photo-dissociation rate & $\kappa$ & $4.2\times10^{-11}$ s$^{-1}$ & 7\\
Self-shielding function & $\Phi$ & $\frac{1-\omega}{(1+x/b_5)^2}+\frac{\omega}{(1+x)^{1/2}}\exp\left[-8.5\times10^{-4}(1+x)^{1/2}\right]$ & 7,8\\
 & & $x=N_{\rm H_2}/5\times10^{14}$ cm$^{-2}$, $\omega=0.035$, $b_5=2$ & \\
Cross-section at $1000$ \AA & $\sigma_d$& $2\times10^{-21}\,(Z /\rm Z_\odot)$ cm$^{2}$ & 7\\
\hline
\end{tabular}
\begin{flushleft}
References: \\
1: \citet{Gnedin+09}
2: \citet{Christensen+12}, 
3: \citet{Cazaux+09},
4: \citet{Burke+83}, 
5: \citet{Tielens+85},
6: \citet{Cazaux+04},
7: \citet{Draine+96}
8: \citet{Sternberg+14}
\end{flushleft}
\caption{\label{table:pararameters}Parameters of the non-equilibrium model for the H$_2$ abundance.}
\end{table*}

\subsubsection{Accounting for unresolved structures}
Cosmological simulations of galaxy formation are limited in spatial resolution,
and even the highest resolution runs employ computational elements that extend for
only a few tens of parsecs. Observations and numerical studies of the turbulent ISM,
on the other hand, reveal a complex gas density distribution on much smaller scales, 
consisting of filamentary structures and clumps \citep[e.g.][]{Glover+07a}. This structure 
is normally approximated in cosmological simulations by introducing a density clumping 
factor, $C_\rho$. However, as already noted by \citet{Micic+12}, this does not take into 
account the full distribution of sub-grid densities, nor the effective density-temperature 
relation, both of which may modify the sub-grid H$_2$ formation and destruction rates. 

Motivated by observations of the GMC density distribution 
\citep[e.g.][]{Kainulainen+09,Schneider+13}, we assume that sub-grid clumps follow a 
lognormal (mass-weighted) probability density function (PDF):
\begin{equation}\label{eq:pdf}
  \mathcal{P}_{\rm M}\,{\rm d}n_{\rm H}=\frac{1}{\sqrt{2\pi}\sigma n_{\rm H}}\,e^{-\frac{(\ln n_{\rm H}-\mu)^2}{2\sigma^2}}\,{\rm d}n_{\rm H}\,,
\end{equation}
where $\mu$ and $\sigma$ are parameters that can determined once a clumping factor 
has been chosen. To do so, note that the average hydrogen density within a computational volume 
element, $\langle n_{\rm H}\rangle$, is simply the integral over the volume-weighted sub-grid 
density PDF, $\mathcal{P}_{\rm V}$:
\begin{equation}\label{eq:cell_density}
  \langle n_{\rm H}\rangle = \frac{\int_0^{\infty}{\rm d}n_{\rm H}\,n_{\rm H}\,\mathcal{P}_{\rm V}}{\int_0^{\infty}{\rm d}n_{\rm H}\,\mathcal{P}_{\rm V}}=e^{\mu-\sigma^2/2},
\end{equation}
where the last expression derives from the fact that $\mathcal{P}_{\rm V}/\mathcal{P}_{\rm M}=\langle n_{\rm H}\rangle/n_{\rm H}$. 
Similarly, $C_\rho\equiv\langle n_{\rm H}^2\rangle/\langle n_{\rm H}\rangle^2=e^{\sigma^2}$, 
so that the sub-grid density PDF is fully determined by $C_\rho$ and the total hydrogen density 
in a cell.

In principle, $C_\rho$ is a local variable whose value depends  
on the turbulent velocity dispersion of the ISM \citep[e.g.][]{Price+11}.
However, for simplicity, and in order to facilitate comparison with previous 
work, we set $C_\rho=10$ which has been shown to reproduce observed H$_2$ fractions 
in nearby galaxies \citep[e.g.][]{Gnedin+09,Christensen+12}.

We assume that unresolved gas concentrations follow a temperature-density relation that
emerges from simulations of the turbulent ISM 
\citep[see Figure 17 in][]{Glover+07b}. These results suggest that, at the densities 
relevant for efficient H$_2$ formation, gas temperatures rarely
exceed $200$~K. We therefore assume that atomic hydrogen and helium
remain neutral within each cell, and neglect the collisional terms in
equations (\ref{eq_form_H2}) and (\ref{eq_form_HI}). The resulting equations can then
be rewritten as
\begin{align}
& \label{eq_H2_int} \dfrac{{\rm d}\langle n_{\rm H_2}\rangle}{{\rm d}t} = \langle\mathcal{R}_f(T) \, n_{\rm HI} \, n_{\rm H}\rangle-\langle G\, \kappa\, \Phi\, e^{-\tau} \,n_{\rm H_2}\rangle , \\
& \label{eq_HI_int} \dfrac{{\rm d}\langle n_{\rm \HI} \rangle}{{\rm d}t} = -2\dfrac{{\rm d}\langle n_{\rm H_2}\rangle}{{\rm d}t}\,,\\
&\label{eq_closure_H_int} \langle n_\HI\rangle+2\langle n_{\rm H_2}\rangle = \langle n_{\rm H}\rangle\,\,\, .
\end{align}

\subsubsection{Solving the differential equations}
Given our assumptions for the sub-grid density and temperature distributions, 
the above rate equations are exact. However, it is impractical to preserve the
information about the abundance of molecular hydrogen at each sub-grid density
between time steps, and further simplifications are needed. We therefore assume
that, within a computational cell, atomic gas transitions to a fully molecular 
state above a critical (sub-grid) density threshold, $n_c$. With this,
equation (\ref{eq_H2_int}) reduces to
\begin{align}\label{rhs}
 & \dfrac{{\rm d}\langle n_{\rm H_2}\rangle}{{\rm d}t} = \langle n_{\rm H}\rangle \left(\int_{0}^{n_c}\,{\rm d}n\,\mathcal{R}_f[T(n)]\,n\,\mathcal{P}_{\rm M} \right.\\
 & \qquad \qquad \left.- \frac{\langle G\rangle\kappa}{2}\int_{n_c}^{+\infty}\,{\rm d}n\,\Phi(n)e^{-\tau(n)}\,\mathcal{P}_{\rm M}\right)\;.\nonumber
\end{align}
Note that, for a lognormal probability distribution of densities, $n_c$ can be obtained solving for the root of 
\begin{equation}
  \langle n_{\rm H_2}\rangle =\frac{\langle n_{\rm H}\rangle}{4}\left[ 1+{\rm Erf}\left(\frac{\mu-\ln(n_c)}{\sqrt{2}\sigma}\right)\right]\,\,.
  \label{H_2_nc}
\end{equation}
At each time step of the simulation, we solve equation~(\ref{rhs}) in 
relevant cells using a variable step-size, variable order, implicit integrator.
Using this we determine the \textit{total} H$_2$ density per cell, which we advect
with the gas flow at each timestep.

\subsubsection{Estimating optical depths}

In order to evaluate equation (\ref{rhs}), the H
and H$_2$ column density (which are necessary to compute the dust optical depth, $\tau(n)$, and the self-shielding 
function, $\Phi$, respectively) must be specified.
A common choice is to compute column densities (surface densities in case of the KMT model)
by means of a Sobolev-like approximation
\begin{equation}
 N_{\rm H}\approx \langle n_{\rm H} \rangle \times l_{\,\rm sob}\;, 
\qquad \qquad\Sigma_{\rm gas}\approx \langle \rho_{\rm gas} \rangle 
\times l_{\,\rm sob}\; ,
 \label{eq:sigma_lsob}
\end{equation}
where $l_{\,\rm sob}=\langle n_{\rm H}\rangle/|\nabla \langle n_{\rm H}\rangle|$, and
$\rho_{\rm gas}$ is the total gas density.
This is a reasonable approximation in simulations where computational elements are of 
comparable extension to the sizes of GMCs \citep{Gnedin+09,Christensen+12,Kuhlen+12}. However, 
because H$_2$ self-shielding takes place in narrow lines, equation (\ref{eq:sigma_lsob})
may over-estimate the H$_2$ column density by a large factor. For this reason, \citet{Gnedin+11}
compute $N_{\rm H_2}$ using the product between the H$_2$ volume density and a fixed characteristic length, $L_c\simeq C_\rho\,10\,\rm pc$, which is tuned to match observations of the H$_2$ fraction as a function of $N_{\rm H}$.
In our model, we compute the column densities of atomic and molecular hydrogen consistently with equation (\ref{eq:pdf}) by simply assuming that the sub-grid density distribution is plane-parallel. Although this highly symmetric configuration cannot exactly hold in a turbulent medium, it provides a reasonable estimate of the relation between volume and column densities and is often used in the literature of molecular clouds \citep[e.g.][and references therein]{Tielens+85,Sternberg+14}. In this case, the total hydrogen mass associated with sub-grid densities below $n_{\rm H}$ is related to the hydrogen column density:
\begin{equation}
 \int_{0}^{n_{\rm H}} {\rm d}n_{\rm H}^\prime\,\mathcal{P}_{\rm M}(n_{\rm H}^\prime)=\frac{1}{\langle n_{\rm H}\rangle\,\Delta x}\int_{0}^{z}{\rm d}z^\prime\,n_{\rm H}\,\,\,.
\end{equation}
Therefore, the final expressions for the column densities are
\begin{align}
 &\label{eq:column_density_H} N_{\rm H}(n_{\rm H})=\frac{\langle n_{\rm H}\rangle\,\Delta x}{2}\left[1-{\rm Erf}\left(\frac{\mu-\ln(n_{\rm H})}{\sqrt{2}\sigma}\right)\right] \\
 &\label{eq:column_density_H2} N_{\rm H_2}(n_{\rm H})=\frac{\langle n_{\rm H}\rangle\,\Delta x}{4}\left[{\rm Erf}\left(\frac{\mu-\ln(n_c)}{\sqrt{2}\sigma}\right)\right.\\
 &\qquad\qquad\qquad\left.-{\rm Erf}\left(\frac{\mu-\ln(n_{\rm H})}{\sqrt{2}\sigma}\right)\right]\,\,\,.\nonumber
\end{align}
As we will show in Section \ref{sec:int_struct_z_2}, our model accurately reproduces the observed relation between the H$_2$ fraction and the column density of neutral hydrogen without introducing additional free parameters.

\subsubsection{Example solutions}
\begin{figure}
  \includegraphics[scale=0.4]{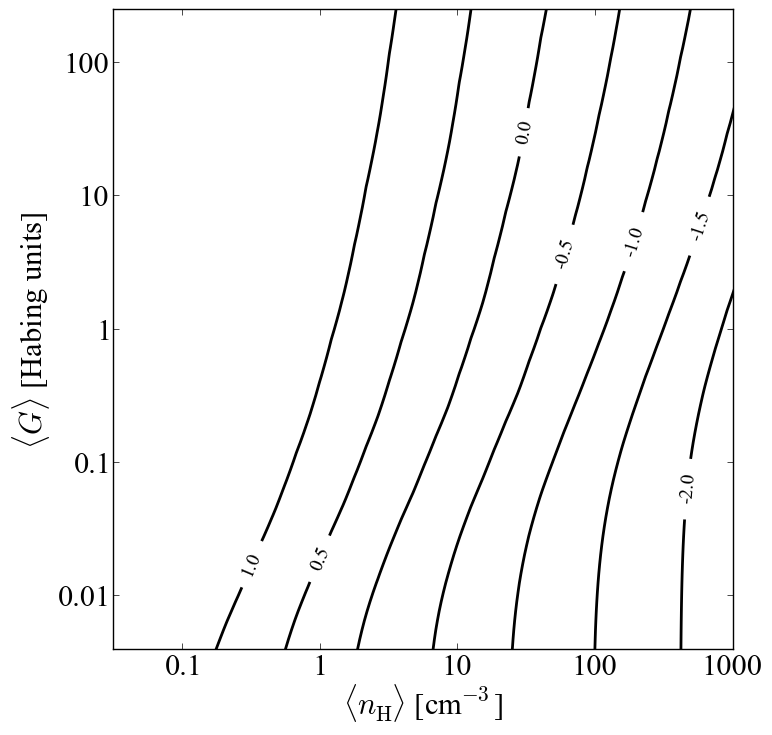}
  \caption{\label{fig:DYN-model}
    Lines of constant $\langle f_{\rm H_2}\rangle=0.5$ as a function of the total hydrogen 
    density, $\langle n_{\rm H}\rangle$, and the interstellar UV field, $\langle G\rangle$, for the DYN model 
    after an integration time of $500$ Myr. Labels along each curve indicate the 
    logarithm of the gas metallicity in solar units.}
\end{figure}

To develop an intuitive understanding of the impact of our sub-grid density distribution 
on H$_2$ formation, we integrate the rate equations for 500 Myr and compute the H$_2$ content 
for cells of fixed density and metallicity in a constant UV field.
The results are shown in Figure \ref{fig:DYN-model}, where curves correspond to 
$\langle f_{\rm H_2}\rangle=2\langle n_{\rm H_2}\rangle/\langle n_{\rm H}\rangle=0.5$ 
in the ($\langle n_{\rm H}\rangle,\langle G\rangle$) plane
for gas metallicities ranging from 10$^{-3}$ to
10 Z$_\odot$. As expected, higher metallicity gas forms H$_2$ more efficiently 
at low densities. This is due to the enhanced rate of H$_2$ formation on dust
grains, but also to the increased optical depth to LW photons (see equations (\ref{eq:column_density_H}) and (\ref{eq:column_density_H2})).
More importantly, note that the 
H$_2$ fraction is nearly independent of $\langle G\rangle$, suggesting that 
our results will not be strongly influenced by our treatment of the UV field.

\begin{figure}
  \includegraphics[scale=0.4]{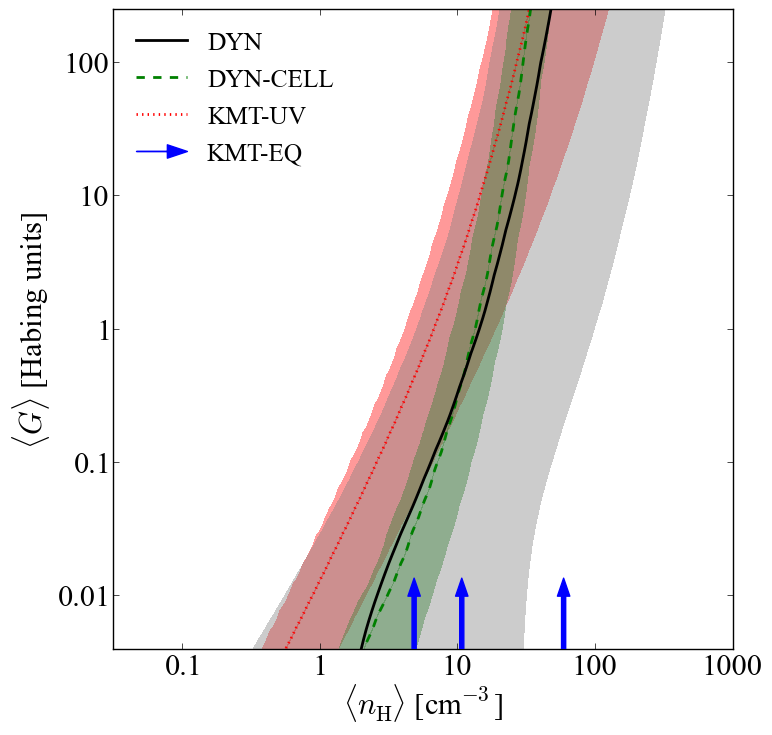}
  \caption{\label{fig:model_comparison} Lines correspond to contours of $\langle f_{\rm H_2}\rangle =0.5$
    for a solar metallicity gas. Different linestyles correspond to the different H$_2$ formation
    models described in Section~\ref{sec:mol_hydrogen_formation}, as indicated in the legend.
    The dashed (green) line, labelled DYN-CELL, shows the results of the DYN model in which sub-grid 
    densities are accounted for using a clumping factor rather than a sub-grid density PDF. 
    Shaded regions of similar colour highlight the zone $0.1 < \langle f_{\rm H_2}\rangle  < 0.9$ for each model.
    Upward pointing arrows mark $\langle f_{\rm H_2}\rangle=0.1,0.5$ and $0.9$ (from left to 
    right) for KMT-EQ, which is independent of $\langle G\rangle$. For the non-equilibrium models, the solution 
    has been computed after a total integration time of $500$~Myr.}
\end{figure}
\subsection{A Comparison of H$_2$ Formation Models}\label{sec:model_comparison}

In Figure~\ref{fig:model_comparison} we compare the predictions of the three H$_2$-formation 
models described above (KMT-EQ, KMT-UV and DYN). Lines show contours of 
$\langle f_{\rm H_2}\rangle=0.5$ in the plane of UV field and gas density for a solar 
metallicity gas; shaded regions indicate the range $0.1<\langle f_{\rm H_2}\rangle<0.9$.
Different line styles correspond to the different models: DYN (solid), KMT-UV (dotted), and 
upward pointing arrows mark $\langle f_{\rm H_2}\rangle=0.1$, 0.5 and 0.9 (from left to right) 
for KMT-EQ (which, by construction, is independent of $\langle G\rangle$). 

We also show, using a dashed (green) line, the time-dependent solution obtained after multiplying 
all quadratic terms by $C_\rho=10$ but without integrating over the sub-grid density PDF. This 
approximation (labelled DYN-CELL) has been used in previous work \citep{Gnedin+09,Christensen+12}, 
and is shown here for comparison. In this case, we set a constant gas temperature of $T=10^3$~K, 
which is typical of star-forming cells in simulations of galaxy formation, and compute the H$_2$ column density as $\langle n_{\rm H_2}\rangle\,L_c$, with $L_c=1$ pc \citep[e.g.][]{Gnedin+11}.The solutions for all 
time-dependent models have been integrated for 500 Myr. Note that within this time, 
a dynamical equilibrium between H$_2$ formation on dust grains and photo-dissociation due
to interstellar radiation has been reached at gas densities
$\langle n_{\rm H}\rangle\simgt1$ cm$^{-3}$ and a longer integration time would only 
impact the very low density regions, without altering our conclusions.

It is worth noting
that the DYN and DYN-CELL models have very similar contours at $\langle f_{\rm H_2}\rangle=0.5$
over roughly five orders of magnitude in UV field. The $\langle f_{\rm H_2}\rangle=0.1$ and 
0.9 contours, on the other hand, differ dramatically. For example, for $\langle G\rangle=1$, the
DYN model predicts an H$_2$ fraction of 0.9 at $\langle n_{\rm H}\rangle\sim 100$ cm$^{-3}$, while in the DYN-CELL model the same H$_2$ fraction is obtained at a density which is approximately 10 times lower. This results from the sub-grid temperature-density relation in 
the DYN model: dense clumps, in this case, are very cold, which inhibits the 
efficient formation of H$_2$ on dust grains. For example, at $\langle n_{\rm H}\rangle\simgt 10^2$ 
cm$^{-3}$ sub-grid clumps have $T\simlt 10^2$~K. Since the H$_2$ formation rate, 
$\mathcal{R}_f$, is proportional to $\sqrt{T}\,S_{\rm H}(T)$, this results in 
lower $\langle f_{\rm H_2}\rangle$ than for the DYN-CELL model, where $T=10^3$~K everywhere.

However, for the same value of $\langle G\rangle$, $\langle f_{\rm H_2}\rangle=0.1$ is reached at 
much lower densities in the DYN model. In this regime, the temperature does not strongly affect the 
H$_2$ formation rate. The difference, in this case, is due to the integration over the sub-grid 
densities, which enhances shielding from LW photons resulting in higher H$_2$ abundances.

At equilibrium, the predictions of the DYN model are consistent with those of the KMT-UV and KMT-EQ models, but the impact of a time-dependent H$_2$ chemistry may be affected by differing treatments of H$_2$ formation on sub-grid scales.
In order to test the impact of these assumptions, we have used each model described above 
to simulate the formation of a massive galaxy at $z=2$. These simulations are described below.

\section{Numerical Simulations}\label{sec:code}
\begin{table*}
\centering
\begin{tabular}{ccccccccc}
\hline
Run & H$_2$ model & $\dot{\rho}_{\rm SF}\propto \langle \rho_i\rangle$ & SNe destroy 
H$_2$&$\ell_{\rm i}$ & $\ell_{\rm f}$ & $\Delta x$ (pc) & $z_{\rm i}$ & $z_{\rm f}$ \\
\hline
STD    & KMT-UV &  gas   & -         & 11 & 17 & 180 & 99 & 2\\ 
DYN    & DYN    &  H$_2$ & - & 11 & 17 & 180 & 99 & 2\\
DYN-SNe  & DYN    &  H$_2$ & \checkmark  & 11 & 17 & 180 & 99 & 2\\ 
KMT-EQ & KMT-EQ &  H$_2$ &-          & 11 & 17 & 180 & 99 & 2\\ 
KMT-UV & KMT-UV &  H$_2$ &-          & 11 & 17 & 180 & 99 & 2\\ 

\hline
\end{tabular}
\caption{Main parameters of the simulations used in this work. Four sub-grid models of increasing complexity have been employed to compute the H$_2$ density. In the simplest case, KMT-EQ,  the chemical-equilibrium abundance of molecular hydrogen is computed assuming that the ISM consists of a cold and a warm phase as in Krumholz et al. (2008). A variant of this technique, in which the flux of LW
photons is computed explicitly by propagating radiation from the stars in the simulation forms the KMT-UV algorithm. Finally, the DYN model returns the non-equilibrium H$_2$ density and takes into account unresolved overdensities as described in Section 2.1.  A slightly modified method is adopted in the DYN-SNe simulation in which SNa feedback destroys H$_2$. In all the simulations, stars are formed proportionally to the density of molecular hydrogen with the exception of the STD run in which SF scales as the density of cold gas. The meaning of symbols is as follows: $\ell_{\rm i}$ and $\ell_{\rm f}$ denote the maximum level of refinement reached at the initial and final redshifts $z_{\rm i}$ and $z_{\rm f}$ while $\Delta x$ gives the linear size of the highest-resolution elements found at
$z_{\rm f}$ in proper units.}
\label{table:simulations_assumptions}
\end{table*}

\subsection{Simulation setup}

We ran several cosmological simulations of the formation of a massive
($\sim 10^{12} \, \rm M_\odot$) galaxy up to redshift $z_{\rm f}=2$ 
using the fully Eulerian code {\sevensize{RAMSES}} \citep{Teyssier+02}. 
Each simulation started from the same initial conditions but employed
different models for H$_2$ and SF, as detailed below (see Table
\ref{table:simulations_assumptions} for a compact summary of our runs).

\subsubsection{Cosmological model}

Each run adopted a flat $\Lambda$CDM cosmological model consistent 
with the \textit{WMAP} 7-year data release \citep{Komatsu+11}. The
corresponding parameters are: $\Omega_{\rm m}=0.2726$, 
$\Omega_{\rm b}=0.0456$, $\sigma_8=0.809$, $n_{\rm s}=0.963$ 
and $h=0.704$. Here $\Omega_{\rm m}$ and $\Omega_{\rm b}$ denote the current 
density parameters for the total matter and the baryonic component, 
respectively;
$\sigma_8$ is the rms mass fluctuation in $8 \, h^{-1}$ Mpc spheres,
linearly extrapolated to $z=0$; $n_s$ is the spectral index of the primordial 
density fluctuation spectrum, and $h$ is the Hubble parameter expressed in units 
of $100 \, {\rm km}\, {\rm s}^{-1}\, {\rm Mpc^{-1}}$.

\subsubsection{Initial conditions}

Initial conditions for our simulations were generated with the {\sevensize{MUSIC}} 
code \citep{Hahn+11} in the following way. We first ran a collisionless
`parent' simulation of a $71$ Mpc box from $z_{\rm i}=99$ to $z_{\rm f}=0$ with 
a spatial resolution of 280 kpc. From the $z=2$ output, we randomly selected
a DM halo with an approximate virial mass\footnote{We define the halo virial mass, 
  $M_{200}$, as that within a sphere of radius $r_{200}$ that encloses a mean density 
  equal to 200 times the critical density for closure, $\rho_{\rm c}=3H_0^2/8\pi G_{\rm N}$
  (where $G_{\rm N}$ is the gravitational constant).} $M_{200}\sim 10^{12}\,\rm M_\odot$ that also had 
a quiescent late-time accretion history. All particles within 3$\,\times\, r_{200}$ (at $z=0$, 
when the halo has a mass $M_{200}\simeq 4\times 10^{12}\,\rm M_\odot$)  were then traced back to the
unperturbed linear density field and the comoving volume enclosing these particles 
($\sim (10\,{\rm Mpc})^3$) was resampled at higher resolution in both DM and gas. The 
global setup includes several nested levels of refinement and periodic boundary conditions.

For our suite of simulations, Lagrangian volume elements within the 
high-resolution region have a length-scale of $\sim 34.6$ comoving kpc. 
This is equivalent to resampling the entire initial $71$ Mpc simulation volume 
with 2048$^3$ cells ($2^{3\ell_{\rm i}}$ with $\ell_{\rm i}=11$). 
For our cosmological parameters and box size, the DM particle 
mass in these runs is $m_{\rm DM}= 1.3 \times 10^6 \,\rm M_\odot$.

\subsubsection{Numerical evolution}

Each simulation was run using a version of the {\sevensize{RAMSES}} code that was modified to 
include the various treatments of H$_2$ physics described in Section~\ref{sec:mol_hydrogen_formation}, 
as well as a new SF routine. {\sevensize{RAMSES}} is an Adaptive Mesh Refinement (AMR) code which uses a second-order 
Godunov scheme to solve the hydrodynamic equations, while trajectories of stellar and DM
particles are computed using a multi-grid Particle-Mesh solver. Between redshifts $z=9$ and 2, we 
output 140 simulation snapshots, equally spaced in 20 Myr intervals.

The AMR technique superimposes finer sub-grids on to the multilevel mesh used to generate the initial 
conditions, resulting in finer resolution in high-density regions. We employ a refinement strategy based 
on the standard `quasi-Lagrangian' criterion: a cell is split if it contains more than eight DM
particles or a baryonic mass greater than $8 \,m_{\rm DM} \,\Omega_{\rm b}/(\Omega_{\rm m}-\Omega_{\rm b})$.
To prevent catastrophic refinement, we enforce a constant physical resolution and match the maximum AMR-level 
to that attained in a pure DM-run, as discussed in section A8 of \citet{Scannapieco+12}. This results in six 
additional levels of refinement before $z_{\rm f}$, corresponding to a maximum level of $\ell_{\rm f}=17$, and to a spatial resolution of 180 pc
at $z=2$.

For the gas component, we assume an equation of state with polytropic index $\gamma=5/3$ and, 
to avoid spurious fragmentation, add thermal pressure using:
$\langle T\rangle=T_{\rm J}\,(\langle n_{\rm H}\rangle/n_{\rm J})^{\gamma-1}$. Requiring the
Jeans length to be resolved with at least four resolution elements \citep{Truelove+97} one finds 
$T_{\rm J}\simeq 2500\,(\Delta x/180~{\rm pc})^{2/3}$ K and $n_{\rm J}\simeq 3.8\,(\Delta x/180~{\rm pc})^{-4/3}$ cm$^{-3}$
for the Jeans temperature and density, where $\Delta x$ is the (physical) length of the resolution 
element \citep{Teyssier+10}.

Our runs include SF, supernova (SNa) feedback and associated metal enrichment, as well as 
cooling from H, He and metals. We adopt the uniform cosmic UV background of \citet{Haardt+12} 
and approximate self-shielding of dense gas by exponentially suppressing it in cells where the gas density 
exceeds $\langle n_{\rm H}\rangle \sim 0.014$ cm$^{-3}$ \citep{Tajiri+98}. In addition, we approximately 
account for interstellar LW radiation in order to solve for the abundance of molecular hydrogen 
(see Appendix \ref{sec:uv}, for further details).

\subsection{Star Formation}\label{sec:star_formation}

Cosmological simulations of galaxy formation lack the spatial resolution required to model the 
cold ISM. SF is therefore implemented stochastically by converting gas mass elements 
into star particles provided that certain physical conditions are satisfied. The prevailing 
approach is to relate the SFR density, $\dot{\rho}_{\rm SF}$, to the total local 
gas density in a cell, $\langle \rho_{\rm gas}\rangle$ and a suitable time-scale, $t_*$, over which 
SF is expected to take place \citep{Schmidt+59}. One common prescription is given by
\begin{equation}
  \dot{\rho}_{\rm SF}=\varepsilon \, \frac{\langle \rho_{\rm gas}\rangle}{t_{*}},
  \label{eq:SFR}
\end{equation}
where $t_*$ is the free-fall time of the gas, 
$t_{\rm ff}=\sqrt{3\, \pi/(32\, G_{\rm N} \, \langle\rho_{\rm gas}\rangle)}$
and $\varepsilon$ is an efficiency parameter.

Our `standard' run (STD) adopts this SF law only for
convergent flows, within cells above a critical
density, $\langle n_*\rangle$, and below a temperature threshold, $T_c$. We set $T_c=10^4$ K, and tune 
$\varepsilon$ and $\langle n_*\rangle$ in order to match the observed KS relation. This gives
$\langle n_*\rangle=2/3~n_{\rm J}$ and $\varepsilon=0.05$ which
corresponds to a SF density threshold of $\langle n_*\rangle\simeq 2.5$ cm$^{-3}$.

Alternatively, SF can be linked directly to the local density of 
molecular hydrogen \citep{Pelupessy+06,Robertson+08,Gnedin+09,Gnedin+10,Feldmann+12, Zemp+12,Christensen+12,Kuhlen+12,Kuhlen+13, Munshi+13, Jaacks+13, Munshi+14, Thompson+14}:
\begin{equation}
  \dot{\rho}_{\rm SF} = \varepsilon \, \frac{\langle \rho_{\rm H_2}\rangle}
{t_{*}}\;.
  \label{eq:mol-SFR}
\end{equation}
Note that this does not require a density threshold, which arises naturally in the equations 
regulating the abundance of H$_2$ (see Figure \ref{fig:DYN-model}): low-density regions, where 
H$_2$ formation is inefficient, are ineligible to form stars. Assuming that SF takes 
place exclusively within GMCs, whose typical densities are of the order of $100$ cm$^{-3}$, we set 
$t_{*}$ in equation~(\ref{eq:mol-SFR}) to be the minimum of the free-fall time-scales computed at the 
cell density and at $100$ cm$^{-3}$ \citep[see also,][]{Gnedin+09}, and adopt $\varepsilon=0.05$, 
as assumed for the STD run. Finally, in the DYN model, we only allow SF in convergent
flows if the 
temperature of a cell is below $10^4$ K.

\subsection{Feedback, metal enrichment and molecules}\label{sec:feedback}

Massive stars end their lives as Type II supernovae (SNe) which inject metals and energy into 
the ISM. Our runs adopt a stellar metallicity yield and an SNa return fraction consistent with a 
\citet{Kroupa+01} initial mass function. We assume that each massive star releases 10$^{50}$ erg 
M$_\odot^{-1}$ of thermal energy into the ISM $10$ Myr after their creation. To approximate the 
adiabatic expansion that follows a SNa explosion, we turn off gas cooling for the next $40$ Myr 
in the affected cells \citep[e.g.][]{Stinson+06,Agertz+12}.

The impact of SNe on the distribution of H$_2$ on $\sim100$ pc scales is far from certain, and 
current simulations of the ISM reach conflicting results \citep[][and references therein]{Krumholz+14}. 
Some authors find that H$_2$ is almost 
completely destroyed  (Walch, personal communication) while others find that it quickly 
reforms due to the short cooling times of the densest regions \citep{Rogers+13}. 
We have bracketed this uncertainty by considering two extreme cases in our DYN model. 
In one, we leave the H$_2$ distribution computed according to our non-equilibrium
model unchanged in cells that have experienced recent SNa events. In a second simulation 
we set the H$_2$ fraction to zero in cells that are 
directly influenced by SNe (i.e. those where the cooling is switched off) and we will refer to this run as DYN-SNe. No 
H$_2$ destruction due to SNe has been 
considered in the simulations based on the KMT models; in this case, H$_2$ is continuously `painted'
on the gas with no memory of the past conditions.

Finally, note that all our models for the formation of molecular hydrogen require the presence of 
dust (and hence metals) in order to catalyse the initial reactions. In fact, we do not follow H$_2$ 
formation in the gas phase which is important only in very small objects (unresolved in our 
simulations) at early cosmic epochs \citep{Abel+97,Abel+98}. We therefore begin our 
simulations from pristine gas assuming 
an SF law given by equation~(\ref{eq:SFR}). These runs are stopped at $z=9$, at which 
point we introduce a metallicity floor of $Z_{\rm floor}=10^{-3}$ Z$_\odot$ in the regions that 
are uncontaminated by prior SF. The $z=9$ outputs are then used as 
the initial conditions for the runs that follow the distribution of H$_2$ molecules from $z=9$ to 
$z=2$. This procedure ensures that high-density, star-forming clumps are enriched with metals more 
efficiently at early times, and also approximately accounts for the enrichment expected from 
unresolved Population III SF \citep[see, e.g.,][for more detail]{Wolfe+05,Wise+12}.

\subsection{Identification of galaxies and haloes}
\label{sec:halofinding}

In all simulation outputs, we identify gravitationally bound objects 
using the Amiga halo finder \citep[{\sevensize{AHF}},][]{Gill+04,Knollmann+09}.
Among other quantities, {\sevensize{AHF}} returns the centre of each halo,
its virial mass, $M_{200}$, and corresponding virial radius, $r_{200}$. 
(Note that halo masses and radii are computed using all matter.)
Within each halo, we define the stellar and gas mass of the
central galaxy as that enclosed within a radius $r_{\rm gal}=0.1\,r_{200}$ 
\citep[e.g.][]{Scannapieco+12}, which gives $r_{\rm gal}\simeq 12.6$ (physical) kpc 
for the largest galaxy in the high-resolution region of our simulations at $z=2$.

In order to link haloes between two consecutive outputs $z_{i}<z_{i-1}$, 
we consider a `descendant' halo identified at $z_{i}$, and search for all of its
`progenitor' haloes at $z_{i-1}$. Progenitors are defined as haloes that have 
DM particles in common with the descendant, and the halo that provides 
most of the mass is referred to as the `main progenitor'.
The history of the main galaxy is tracked by studying the evolution
of the material within 10 per cent of the virial radius of the main progenitor, which we compute
in each simulation output.

\section{Results}\label{sec:analysis}

\begin{table*}
\centering
\begin{tabular}{lcccccccccc}
\hline
Model          &  $M_*$ & $M_{\rm gas}^{\rm cold}$ & $M_{\rm gas}^{\rm hot}$ & $M_{\rm H_2}$ & SFR & $\langle Z\rangle_{\rm M}$ & $\langle Z\rangle_*$ & $\langle n_{\rm H}\rangle_{20}^{\rm SFR}$ & $\langle n_{\rm H}\rangle_{50}^{\rm SFR}$ & $\langle n_{\rm H}\rangle_{80}^{\rm SFR}$\\
                     & ($10^{11}$ M$_\odot$)& ($10^{11}$ M$_\odot$)& ($10^{11}$ M$_\odot$)& ($10^{11}$ M$_\odot$)& (M$_\odot$ yr$^{-1}$) & (Z$_\odot$) & (Z$_\odot$) & (cm$^{-3}$) & (cm$^{-3}$) &(cm$^{-3}$) \\
\hline
STD     & 1.15 &  0.35 &  0.27 &  0.36  &  59.6  & 0.92  & 0.72 & 10.2 & 29.7 & 83.7 \\
DYN     & 0.77 &  0.31 &  0.53 &  0.44  &  42.6  & 0.72  & 0.49 & 19.9 & 71.3 & 178.3 \\
DYN-SNe & 0.58 &  0.44 &  0.56 &  0.39  &  40.8  & 0.58  & 0.41 & 18.2 & 68.3 & 224.9 \\
KMT-UV  & 0.74 &  0.43 &  0.42 &  0.39  &  58.4  & 0.72  & 0.51 & 9.2  & 21.1 & 52.7 \\
KMT-EQ  & 1.36 &  0.27 &  0.06 &  0.12  &  67.0  & 1.07  & 0.82 & 4.8  & 10.4 & 23.7 \\

\hline
\end{tabular}
\caption{Physical properties for the main galaxy (i.e. measured within $r_{\rm gal}=0.1\,r_{200}$) at $z=2$. The columns indicate the stellar, the 
  cold ($T<10^4$ K), the hot ($T>10^4$ K) and the molecular gas mass, the SFR, the mass-weighted 
  gas and stellar metallicities, and 20th, 50th and 80th percentiles of the SFR-weighted gas density.}
\label{table:par_r200}
\end{table*}

\subsection{Mass assembly history}

\begin{figure}
  \includegraphics[scale=0.4]{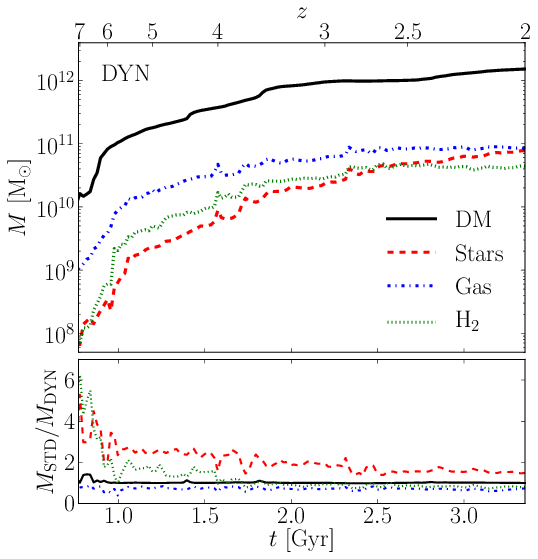}
  \caption{\label{fig:mass_profiles} Assembly history of the main galaxy and its DM
    halo in the DYN run. The solid (black) line shows the evolution of the DM mass within $r_{200}$;
    the mass of stars (red dashed lines), gas (blue 
    dot-dashed lines) and molecular hydrogen (green dotted lines) measured within $r_{\rm gal}$
    are also shown. The bottom panel shows the ratio between the mass (of each component) measured in the STD simulation to that in the DYN one.}
\end{figure}

In the top panel of Figure \ref{fig:mass_profiles} we show the mass build-up of the main galaxy
in different components for the DYN run over the redshift range $z=7$-2.
In the bottom panel, instead, we plot the same quantities for the STD run, but normalized to the masses in the DYN simulation.
The mass-assembly history of the main halo (thick black line) is virtually identical
for the two models: the halo has a final DM mass of $1.8\times10^{12}$ M$_\odot$ 
and exhibits a relatively quiescent recent formation history, with only three major-merging 
events that take place at $z\sim6$, $\sim4.4$ and $\sim3.5$. At $z=2$ the halo contains more 
than 100 substructures with masses above $10^8$ M$_\odot$ (corresponding to $\sim$100 DM particles), 
but more than 50 per cent of its stellar mass, and almost all of the molecular hydrogen, are located within
the central galaxy.

The coloured lines show the time evolution of the masses
of the different components within $r_{\rm gal}$: stars are shown using a dashed (red) line, the
total gas mass as a dot-dashed (blue) line, and the mass of molecular hydrogen 
as a dotted (green) line. Note that, in the STD run, the local H$_2$ abundance is computed in post processing using 
the KMT-UV model.

We can distinguish three main evolutionary epochs, each lasting for
approximately 1 Gyr: one in which the main galaxy is assembled out of several small sub-galactic
objects (prior to $z=5.8$), followed by several minor mergers that take place between $3.6<z<4$ 
and $2.7<z<3.3$, and a final quiescent approach to $z=2$ perturbed by the close passage of a 
relatively massive satellite at $z\sim 2.15$. Note that the stellar mass grows steadily 
with time while the gas and H$_2$ masses remain approximately constant after $z\sim3$.

The main galaxy in the STD model contains more stars than the DYN one
at all epochs. Within the first Gyr, the STD galaxy forms five times more stars while
the difference gradually decreases with time (the ratio reduces to $\sim1.5-2$ at  $t\simeq3$ Gyr).
Early SF is suppressed in the DYN run because H$_2$ is created at a slow pace in a
metal-poor environment. At $z = 6$ ($t=0.9$ Gyr), the mean (mass-weighted) gas
metallicity in the galaxy is $\langle Z\rangle_{\rm M}=0.1$ Z$_\odot$ and the characteristic time-scale 
for H$_2$  formation is as large as
$\tau_{\rm f}=1/(\mathcal{R}_f\,\langle n_{\rm H}\rangle)\propto \langle Z\rangle^{-1}\sim 1$ Gyr (for 
the typical densities in the galaxy). 
In the following 3 Gyr, however, the ISM is steadily polluted with metals and $\langle Z\rangle_{\rm M}$ increases by roughly 
one order of magnitude. This promotes H$_2$ formation at lower gas densities, and subsequently enhances SF.
\begin{figure*}
\includegraphics[scale=0.35]{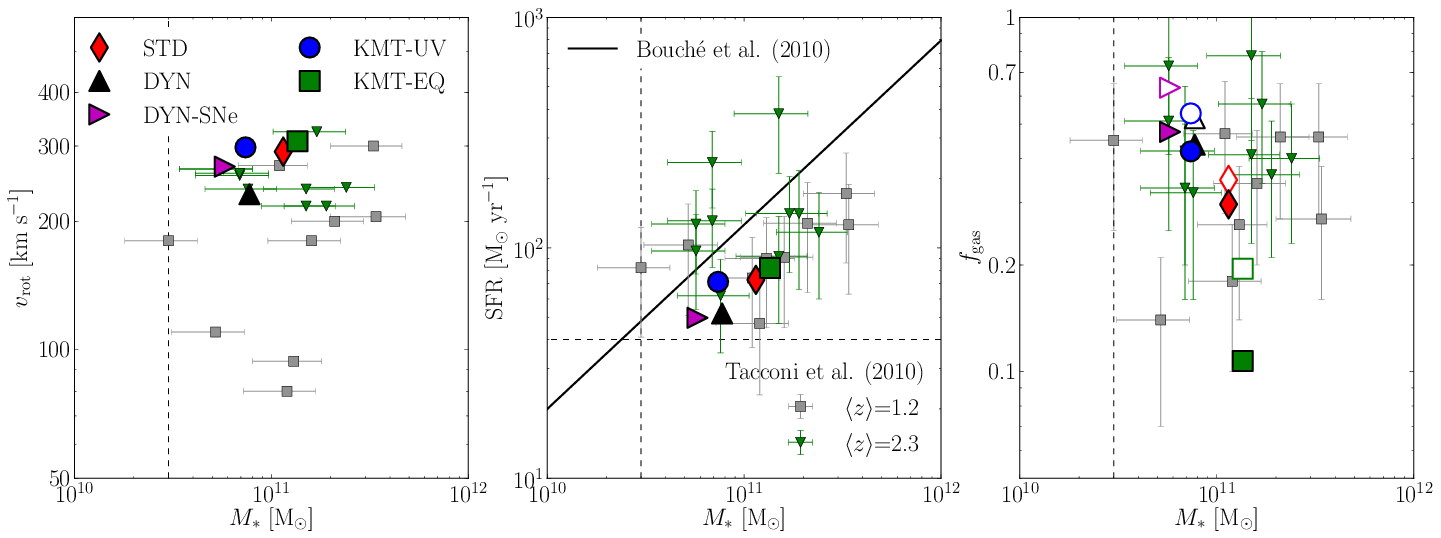}
\caption{\label{fig:tacconi} Gas rotational velocity (left-hand panel), star formation rate
  (middle panel) and gas fraction (right-hand panel) plotted versus the stellar 
  mass of the main galaxy at $z=2$ for all runs. Heavy symbols correspond to our 
  simulated galaxies, while small ones show the observed sample of \citet{Tacconi+10}, which
  have been divided into low (grey squares) and high (green triangles) redshift subsamples. 
  The solid black line in the middle panel is the galaxy SFR-main sequence at $z=2$ from 
  \citet{Bouche+10}. In the right-most panel, open symbols correspond to $f_{\rm gas}$
  measured using the total gas mass, rather than the molecular component alone. The vertical 
  and horizontal dashed lines show the selection criteria in stellar mass ($M_*>3\times10^{10}$ 
  M$_\odot$) and star formation rate (SFR$>40$ M$_\odot$ yr$^{-1}$) for the observational samples 
  of \citet{Tacconi+10}.
}
\end{figure*}
The simulations that are not shown in Figure \ref{fig:mass_profiles} follow similar
patterns to either the DYN or the STD models. For instance, in the KMT-UV and DYN-SNe simulations
SF is also hindered at early times (in particular the DYN-SNe run forms nearly 20 per cent less stars than the DYN one at all times as the overall molecular mass is influenced by SNa explosions). On the other hand, the KMT-EQ run, where the H$_2$ abundance is not directly influenced by
the presence of a dissociating radiation field, does not show any significant difference with respect to 
the STD case.

To facilitate a comparison between the different models, we 
list in Table \ref{table:par_r200} some key properties
of the main galaxies in each simulation at $z=2$, including their total stellar mass, 
$M_*$, as well as the gas masses in hot, cold and molecular components.
We also list the SFR (averaged over the past 20 Myr), the 
average gas and stellar metallicities and the 20th, 50th and 80th percentiles of the 
SFR-weighted gas density ($\langle n_{\rm H}\rangle^{\rm SFR}$). 
(All quantities have been measured within $r_{\rm gal}$.)

Due to their high spatial resolution, all our simulations form stars only at substantially higher gas densities
than in most previous work. The SF threshold we adopt in the STD model is comparable
to that used in the ERIS simulation \citep{Guedes+11} and is several orders of magnitude above
the commonly adopted value of $\sim 0.01-0.1$ cm$^{-3}$. Our DYN and DYN-SNe models form stars at even higher
densities (by more than a factor of 2, see Table \ref{table:par_r200}). On the other hand, in the KMT-EQ and 
KMT-UV runs, stellar particles tend to be created in cells with somewhat lower densities than in the STD run. This 
has important consequences. In fact, when stars form only in correspondence of the highest density peaks,
energy injection from SNe is strongly clustered and creates a very inhomogeneous ISM. Moreover, stellar 
feedback becomes more efficacious and plays a decisive role in shaping the structure of the emerging galaxy.
For instance,  highly peaked SF is more efficient in removing low-angular momentum gas at high 
redshift and thus reduces the mass of the final central bulge component 
\citep[e.g.][]{Robertson+08,Ceverino+10,Brook+11,Brook+12,Guedes+11}.

In addition, in all runs in which SF is regulated by molecules (with the exception of KMT-EQ) the galaxy 
has a lower stellar mass and harbours a large gas reservoir. This is a direct consequence of suppression 
of SF at early times. For instance, in the DYN simulation, the gas reservoir is nearly 40 per 
cent higher than in the STD case.

\subsection{Comparison with observations}
\subsubsection{Global properties of the galaxy at $z=2$}

In Figure \ref{fig:tacconi} we compare several characteristics of our 
simulated galaxies at $z=2$ to observations of star forming galaxies at
high-redshift. Outsized points in the left-hand panel plot the rotational velocity
of the gas, $v_{\rm rot}$, versus stellar mass for our simulations. Smaller
symbols show the data set of \citet{Tacconi+10}, which is divided into a high ($\langle z\rangle=2.3$) 
and a low redshift sample ($\langle z\rangle=1.2$). To make a meaningful comparison with this data --
for which $v_{\rm rot}$ was determined from CO line emission -- 
we compute the rotational velocity using only the cold gas component.

The SFR versus stellar mass is shown in the middle panel. Large and
small symbols have the same meaning as before. All our model galaxies, independent of the assumed
H$_2$ and SF laws, are forming stars at a similar rate, which is 
in good agreement with the observational data set, as well as with the theoretically determined
SF main sequence at $z=2$ given in \citet{Bouche+10}.

Finally, the right-hand panel shows the gas fraction,
defined $f_{\rm gas}=M_{\rm gas}/(M_{\rm gas}+M_*)$, versus galaxy stellar mass.
Observationally, $M_{\rm gas}$ is determined from CO luminosity which
is first converted into a molecular mass and then multiplied by a factor of
1.36 to account for helium atoms that should be well-mixed with
the molecules. This estimate coincides with the actual gas mass only if the 
contribution from atomic hydrogen is negligible. 
In our simulations, this never holds true. Taking the DYN run at $z=2$ as an example, 
Table \ref{table:par_r200} shows that the cold atomic gas (H and He), the hot gas and the molecular hydrogen
have all nearly identical masses.
Therefore, in order to fairly compare the numerical results
against the sample of \citet{Tacconi+10}, we determine $f_{\rm gas}$ using the relationship 
$M_{\rm gas}=1.36\, M_{\rm H_2}$. These are shown as solid points in Figure \ref{fig:tacconi}.
For comparison, we also show, using open symbols, the \textit{total} gas fraction, which includes
both atomic and molecular components.

In general, the rotational velocities, SFRs and gas fraction of our simulated galaxies agree well 
with the observational data set. Likewise, the simulations nicely match measurements of $f_{\rm gas}$
with the possible exception of the KMT-EQ model whose predictions (when
based on H$_2$ alone) fall below the observed data.

\begin{figure}
  \includegraphics[scale=0.4]{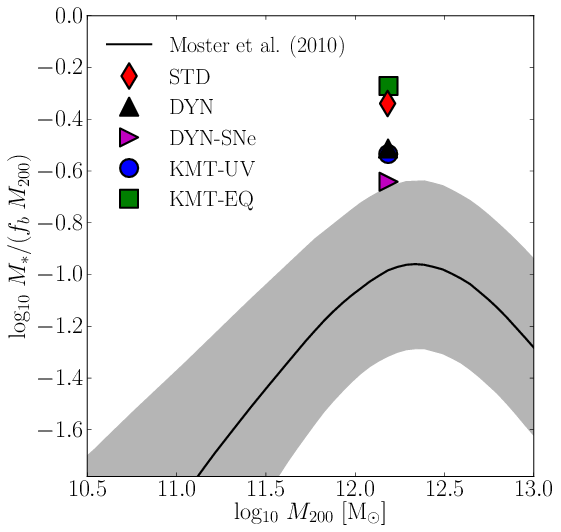}
  \caption{\label{fig:moster_plot} Stellar-to-halo-mass relation at $z=2$ for our simulations (solid symbols) and the observed relation at $z=2$ from \citep[][solid line]{Moster+13}. The grey shades indicate the 1 $\sigma$ confidence level.}
\end{figure}

Another challenging test for the simulated galaxies is to check how their stellar mass-halo mass (SMHM) 
relation compares with the recent semi-empirical constraints which have been derived contrasting the 
galaxy number counts and the number density of DM haloes in $N$-body simulations with the abundance 
matching technique \citep[see][and references therein]{Behroozi+10}. In Figure \ref{fig:moster_plot} we 
plot the SMHM relation at $z=2$ (note that the stellar mass of the central galaxy is normalized to 
$f_b\,M_{\rm 200}$, where $f_b=\Omega_b/\Omega_m$) for our simulations (solid symbols) and compare it 
with the recent observational results at $z=2$ by \citep[][shaded area and solid line]{Moster+13} . The 
STD and KMT-EQ runs are in tension with the data: they produce far too many stars given the mass of 
the DM halo. This is a well-known problem in cosmological simulations in which
no strong feedback mechanisms beyond SNa explosions are included 
\citep[see][]{Scannapieco+12}. However, the simulations in which the H$_2$ abundance is computed 
with an explicit treatment of photo-dissociation behave differently. In fact, molecule-regulated 
SF acts as an effective feedback mechanism and helps reducing the stellar mass of the 
central galaxy. In particular, our DYN-SNe run lies within the 1-$\sigma$ uncertainty of the observed 
SMHM relation. Similar results are found at higher redshifts.

\begin{figure*}
  \includegraphics[scale=0.4]{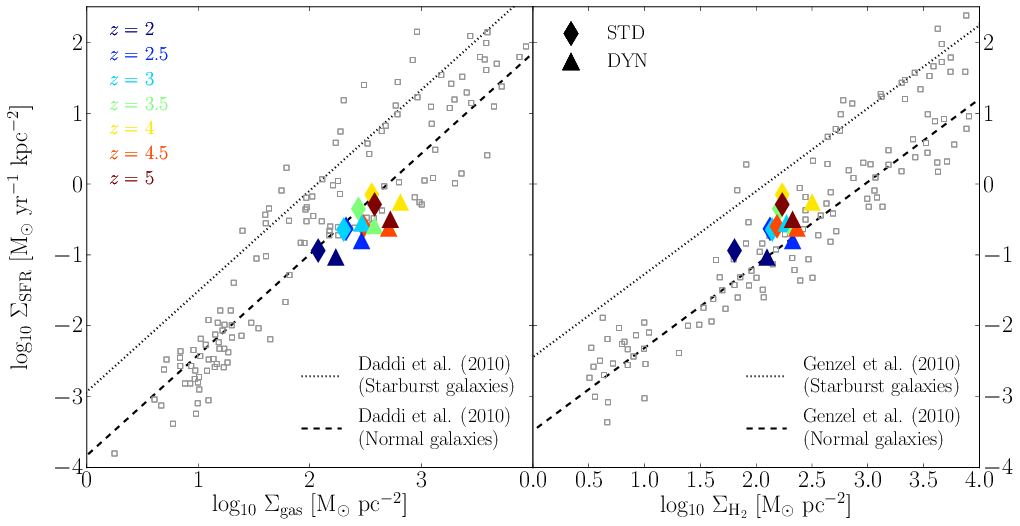}
  \caption{\label{fig:kslaw}The KS relation for the total gas  (left-hand panel) 
    and  for the molecular hydrogen (right-hand panel). Outsized coloured symbols show the
    relation measured for our main galaxy at different redshifts in the STD (diamonds) and DYN (triangles) simulations. We have integrated the gas and young stellar (i.e. 
    $t_s\le20$ Myr) mass distribution over a cylinder of radius and height $r_{\rm gal}$, oriented along
    the $z$-axis of the computational box.
    The dashed and dotted lines show the
    average observed relations for normal/quiescent galaxies and merging/starbursts systems, 
    respectively, and grey squares the observed data for individual galaxies 
    from \citet{Daddi+10} and \citet{Genzel+10}.}
\end{figure*}

\subsubsection{The KS relation}

In this section, we compare how the area-averaged SFR, 
$\Sigma_{\rm SFR}$ and corresponding gas surface densities, $\Sigma_{\rm gas}$ 
and $\Sigma_{{\rm H}_2}$, relate to each other in our simulations.
To do this, we compute all surface densities using a projection of the galaxies over
a cylinder, with radius and height equal to $r_{\rm gal}$, oriented along the $z$-axis
of the computational box.
In order to match observations that measure nebular emission
lines (H$\alpha$, O{\small\rm II}), which are sensitive to the light
of young stars, we average the instantaneous SFR
over the typical lifetime of OB stars, which we take to be 20 Myr. 
Note that we have verified that our results are robust with respect to small changes 
in the orientation, radius and height of the cylinder as well as stellar age.

In the left-hand panel of Figure~\ref{fig:kslaw}  we plot
the KS relation for the total gas, while the right-hand panel
shows the relation for the molecular gas.
Outsized coloured symbols show the results for the main galaxy
at different redshifts, while small grey squares refer to observations
of individual galaxies \citep{Daddi+10,Genzel+10}. In each panel, power-law fits to observations of quiescent and starburst galaxies
are shown with a dashed and dotted line, respectively.

By construction, the STD simulation perfectly reproduces the KS relation (expressed in terms of $\Sigma_{\rm gas}$) for normal galaxies as the parameters of the Schmidt law in equation (\ref{eq:SFR}) were tailored to achieve this result. 
The very same value of the efficiency parameter ($\varepsilon=0.05$) has been used in all the other simulations together with equation (\ref{eq:mol-SFR}). We find that all the runs are in good agreement (within the measured scatter) with the observed KS relation for the total as well as for the molecular gas.

\begin{figure*}
  \includegraphics[scale=0.3]{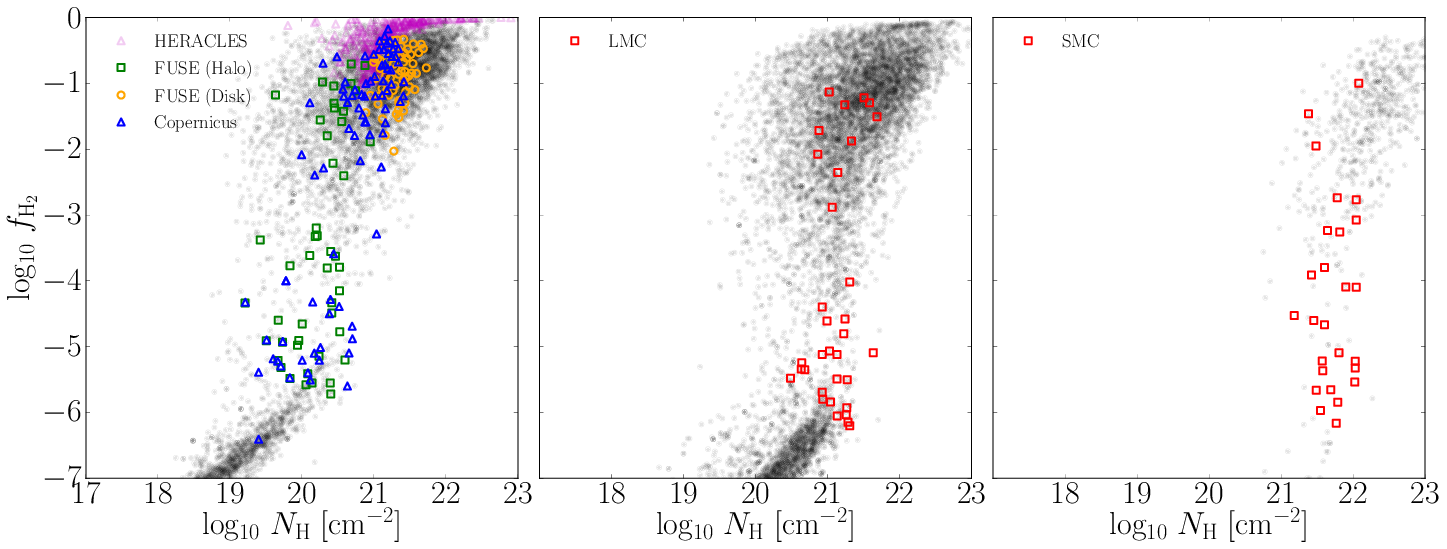}
  \caption{\label{fig:h2_transition} Total hydrogen column density, $N_{\rm H}=N_\HI+2\,N_{\rm H_2}$, versus the average H$_2$ fraction, $2\,N_{\rm H_2}/N_{\rm H}$, measured in a position-position-velocity data cube of 180 pc $\times$ 180 pc $\times$ 40 km s$^{-1}$ for the DYN simulation. Black points in the left-hand panel have been selected at $z\sim2$ and have, within a scatter of 30 per cent, $\bar{Z}=$ Z$_\odot$ and $\bar{G}=1$. Coloured symbols show observations of the H$_2$/\HI~transition in the MW: blue triangles are taken from the Copernicus Survey \citep{Savage+77}, while orange circles and green squares from the FUSE disc \citep{Shull+04} and halo survey \citep{Gillmon+06}, respectively. Purple points show observations from the HERACLES survey for a sample of 30 nearby galaxies \citep{Schruba+11}. Points in the middle panel are selected at $z\sim3.5$ and have mean metallicity and UV field (i.e. $\bar{Z}=0.3$ Z$_\odot$ and $\bar{G}=10$) comparable to the LMC \citep{Tumlinson+02}.  Points in the right-hand 
panel have been selected at $z\sim5.5$ and have similar physical conditions (i.e. $\bar{Z}=0.1$ Z$_\odot$ and $\bar{G}=100$) to the SMC \citep{Leroy+07}.}
\end{figure*}

\begin{figure*}
  \includegraphics[scale=0.35]{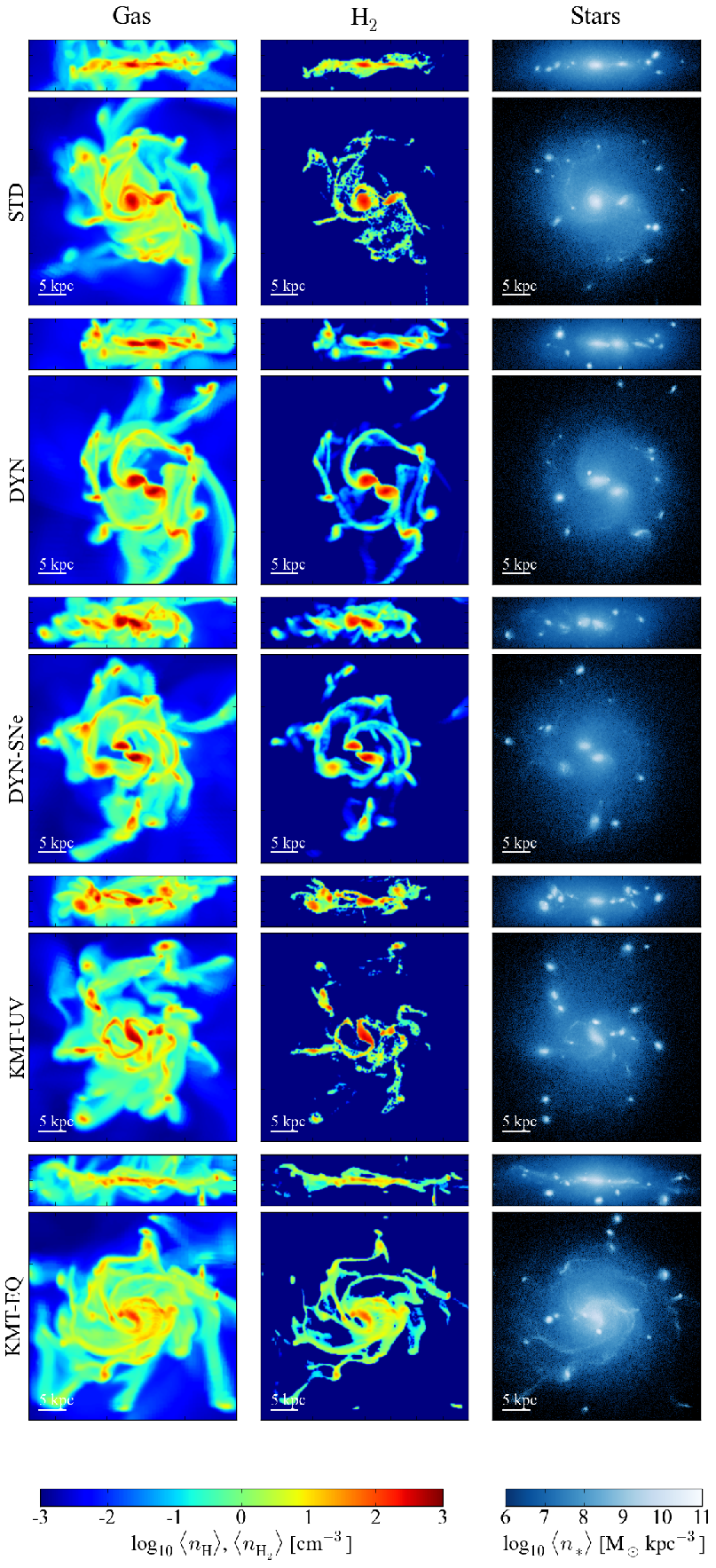}
  \caption{\label{fig:projections} Face-on and edge-on views of the disc of the main galaxy
    for (from top to bottom) the STD, DYN, DYN-SNe KMT-UV and KMT-EQ runs at $z=2$ in a box of $40$ $\times$ $40$ $\times$ $10$ kpc. From 
    left to right, panels show the maximum hydrogen, H$_2$ and stellar densities along the line of sight, respectively.}
\end{figure*}

\subsubsection{The $\rm H_2$/\HI~transition}

Observational studies of the ISM show that the interface between the \HI~and H$_2$-dominated regions is 
rather sharp and occurs around a given hydrogen column density whose value depends on the local metallicity 
and on the intensity of UV radiation. Measurements from absorption spectra of distant quasars and nearby 
stars in the MW \citep{Savage+77,Shull+04,Gillmon+06,Wolfire+08}, the Large Magellanic Cloud 
\citep[LMC][]{Tumlinson+02} and the Small Magellanic Cloud  \citep[SMC][]{Leroy+07} are reported in Figure 
\ref{fig:h2_transition} (coloured symbols). Here, we also plot data extracted from the DYN simulation 
(dots). To be consistent with the observational studies, we compute the H$_2$ and \HI~column densities 
($N_{\rm H_2}$ and $N_\HI$) and the average H$_2$ fraction, $f_{\rm H_2}=2\,N_{\rm H_2}/N_{\rm H}$, 
within a position-position-velocity data cube of 180 pc $\times$ 180 pc $\times$ 40 km s$^{-1}$. In 
order to make a meaningful comparison, we separate data with different metallicity ($\bar{Z}$) and 
interstellar UV field ($\bar{G}$). In the left-hand panel, we show the results extracted from the simulation 
at $z\sim2$ that match the physical conditions of the MW (i.e. $\bar{Z}=$ Z$_\odot$ and $\bar{G}=1$, 
within a scatter of 30 per cent). In addition, we also plot recent measurements from a sample of 30 
nearby galaxies with nearly solar metallicity from the HERACLES survey \citep{Schruba+11}. (Their 
column densities have been computed in tilted rings 15 arcsec wide and corrected for inclination.)

The middle and right-hand panels, instead, refer to the LMC ($z\sim3.5$ with $\bar{Z}=0.3$ Z$_\odot$ and 
$\bar{G}=10$) and the SMC ($z\sim5.5$ with $\bar{Z}=0.1$ Z$_\odot$ and $\bar{G}=100$), respectively. 
In all cases, we find that the DYN model is in excellent agreement with observations. The simulations 
that are not shown in Figure \ref{fig:h2_transition} give similar results. In particular, while the 
\HI~to H$_2$ transition is virtually identical in the DYN-SNe and DYN runs, the simulations based on 
the KMT models correctly locate the value of $N_{\rm H}$ at which $f_{\rm H_2}\sim0.1$ but tend to 
predict too sharp a transition especially at $z\sim2$ when the gas metallicity is high.

\subsection{The internal structure of the galaxies at $z=2$}\label{sec:int_struct_z_2}

\subsubsection{The distribution of molecular hydrogen}\label{sec:molecular_hydrogen_properties}

Despite the fact that all models agree reasonably well with global 
observations of high-redshift galaxies, the detailed properties 
of the stellar, molecular and gaseous components in each simulation are
rather different. For example, in Figure~\ref{fig:projections} we show the 
total hydrogen (left), the molecular hydrogen
(centre) and the stellar particle (right) distributions at $z=2$ for the main 
galaxy in all of our simulations.
The panels show from top to bottom the results for the STD, DYN, DYN-SNe, KMT-UV and KMT-EQ runs, respectively.

The different assumptions 
used in these H$_2$-formation models are readily apparent in the H$_2$ 
density distributions. Notice how including a sub-grid model for 
GMCs results in more molecular hydrogen in low-density gas. Nevertheless, the 
low densities in the outer regions contribute very little
to the total H$_2$-mass budget in the DYN model.

\begin{figure}
  \includegraphics[scale=0.4]{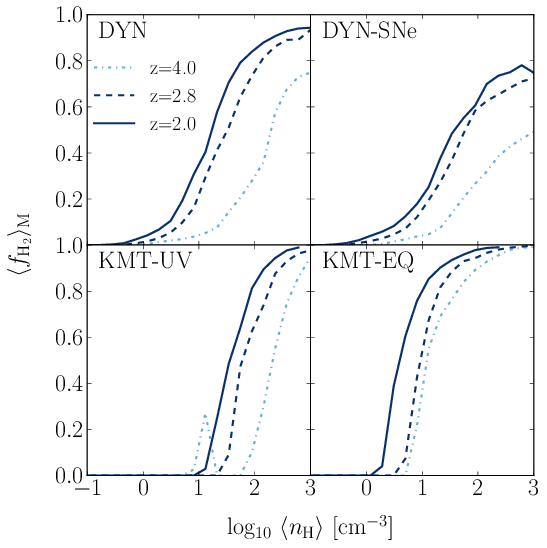}
  \caption{\label{fig:evH2} Distribution of the mass-weighted H$_2$ fraction as a function 
    of cell density for the models with molecule-regulated SF. Different line styles 
    show results at three redshifts, as indicated in the legend.}
\end{figure}

A more quantitative analysis is provided in Figure~\ref{fig:evH2}, 
where we show the time evolution of the mean mass-weighted H$_2$ 
fraction, $\langle f_{\rm H_2}\rangle_{\rm M}$, as a function of the cell density for all of our simulations.
We focus on three different epochs: two merging events ($z\sim 4$ and 2.8) and the final output at $z=2$.

All models show a sharp H$_2$/\HI~transition which progresses towards lower 
densities with time. In the KMT models this is expected since atomic gas is converted into molecules 
only if its surface density is greater than a critical value that
is inversely proportional to the local metallicity. Similarly, for the time-dependent models,
higher metallicity gas produces H$_2$ more rapidly.
Therefore, the shift of the transition 
scale reflects the increasing metal content of the gas.

Finally, the DYN run shows two peculiarities: first, an appreciable H$_2$ 
fraction is found at much lower mean densities than in the KMT models, and second, 
$\langle f_{\rm H_2} \rangle_{\rm M}$ stays well below unity, even in the densest regions at early times.
The first of these is due to the sub-grid density PDF adopted in the DYN model: 
the mean density of a cell is not representative of the regions in which H$_2$ is 
formed. The latter is due to the effect of a time-dependent H$_2$ formation model, 
which slowly gets to equilibrium as the metallicity of the galaxy increases with time.
Note also that, when H$_2$ destruction from SNa events is included, the DYN run predicts 
less molecular hydrogen in high-density regions. For instance, one can see in the top-right 
panel of Figure \ref{fig:evH2} that the DYN-SNe run shows $\sim$20 per cent less molecular 
hydrogen at $\langle n_{\rm H}\rangle>$ 10 cm$^{-3}$.

A key assumption of the H$_2$-models that we have considered here is
that the abundance of dust traces the gas metallicity. Since H$_2$ forms rapidly in dust-rich 
environments, this hypothesis has important consequences. In fact, at all epochs and in all 
simulations, we measure a tight correlation between the mass in metals, $M_Z$, 
and the H$_2$ mass, $M_{\rm H_2}$, at the cell level. 
For example, in the DYN model, the relation between  $M_Z$ and $M_{\rm H_2}$ can be accurately 
described by a power law; the linear correlation coefficient in the $\log \, - \, \log$ plane is 
$\sim 0.99$, and similar values are found for the KMT models, above a threshold mass of metals, 
$M_Z\simgt 2\times10^3$ M$_\odot$.  The best-fitting power-law parameters, however, evolve slowly with 
redshift. For example, in the DYN run, we find  that $M_{\rm H_2}\propto M_Z^\alpha$ where 
$\alpha$ grows smoothly from 0.96 to 1.07 in the redshift interval $5<z<2$.
The constant of proportionality is $\sim$30 when masses are measured in units of M$_\odot$. 
After performing a careful calibration against the parameters of the 
sub-grid models, this tight correlation could be exploited to run computationally inexpensive 
simulations that link the abundance of molecular hydrogen directly to that of the metals.

\subsection{Build-up of the galaxy and its satellites}

\begin{table}
\centering
\begin{tabular}{l|cc}
\hline
Model & $M_*(r<1\,\rm kpc)$   & $t_s(r<1\,\rm kpc)$\\
      & ($10^{10}$ M$_\odot$) & (Gyr) \\
\hline
STD      & 1.1 & 1.1 \\
DYN      & 0.4 & 0.6 \\
DYN-SNe  & 0.3 & 0.5 \\
KMT-UV   & 0.4 & 0.7 \\
KMT-EQ   & 0.9 & 1.3 \\

\hline
\end{tabular}
\caption{\label{table:stars_1kpc} Stellar mass and mean stellar age within 1 kpc from the galaxy at $z=2$ for all the runs.}
\end{table}

\label{sec:stellar_disc}
\begin{figure*}
  \includegraphics[scale=0.32]{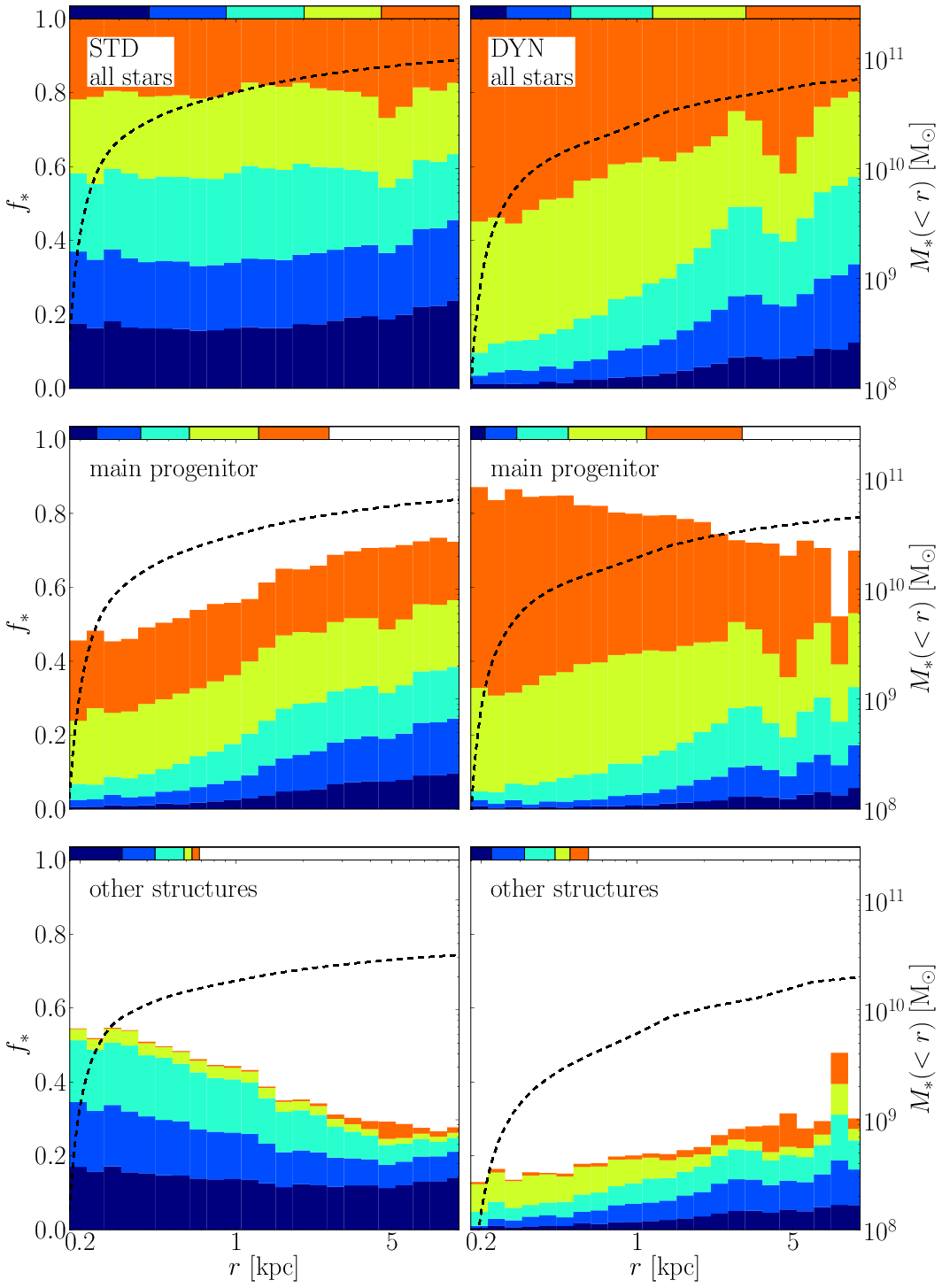}
  \caption{\label{fig:rad_distr_stars} The radial distribution of stars within the galaxy at $z=2$ for the 
    STD (left-hand panels) and DYN (right-hand panels) simulations.
    The first row shows the distribution for all the stars in the galaxy, the second one for the stars 
    formed in the most massive progenitor of the host halo at $z=2$, the third one for the stars formed in other structures.
    Different coloured regions indicate the fraction of stars as a function of radius sorted
    in five stellar age bins: $[3.35,1.80],\,[1.80,1.39],\,[1.39,0.98],\,[0.98,0.48],\,[0.48,0]$ 
    Gyr (colour-coded so that dark-blue shades corresponds to earlier epochs). 
    These bins are chosen so that, in the STD run, 
    the stellar mass produced in each of them is equal to one fifth of the galaxy stellar mass 
    at $z=2$. The coloured bars on top of each panel are scaled in size to match the fraction
    of stellar mass that is formed within each time bin. The cumulative stellar mass distribution 
    (dashed lines) is also plotted in each panel.
    }
\end{figure*}
\begin{figure*}
  \includegraphics[scale=0.5]{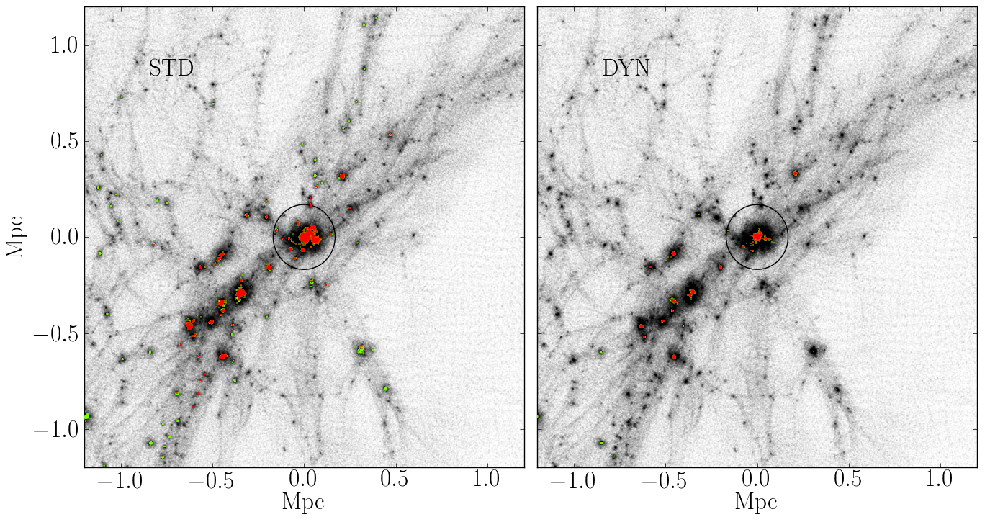} 
  \caption{\label{fig:snapshot_z_5} Comoving spatial distribution (in a 1.2 Mpc thick slice projected 
    along one axis of the simulation box) of the DM (black points) and of the stars (green 
    points) at $z = 5.5$ for the STD (left-hand panel) and DYN (right-hand panel) runs. Stars 
    that are found in the main galaxy at $z=2$ are shown in red. The plots are centred on the main 
    halo, indicated with a black circle of radius $r_{\rm 200}$.}
\end{figure*}

The evolution of the H$_2$ (or total gas) profile in our simulations
determines the structure of the stellar disc. In order to better assess
the differences between our runs, we study the radial distribution of
stars at $z=2$ within the galaxy in bins of stellar age.
We choose the bins so that, in the STD run, the stellar mass produced
in each is equal to one fifth of the galaxy's total stellar mass at $z=2$ 
(corresponding to age bins of [3.35,1.80], [1.80,1.39], [1.39,0.98], [0.98,0.48], [0.48,0] Gyr).
The results for the STD (left-hand panels)
and DYN (right-hand panels) runs are plotted in the top row of Figure
\ref{fig:rad_distr_stars} (qualitatively similar results are obtained when using the DYN-SNe model instead of DYN).
One can see that the DYN model shows significant differences
in the age distribution of stars, especially at early formation epochs
(represented with darker
shades of blue) and in the innermost $5$ kpc. These regions are
populated by younger stars in DYN than
in the STD case. For instance, the mean age of the stellar particles
located at $r<1$ kpc
is 1.1 Gyr in the STD simulation, but only 0.6 Gyr in DYN.
The total stellar mass within 1 kpc from the galaxy centre at $z=2$
is also different. In fact, we find values of $M_*(r<\, 1\, {\rm
kpc})=1.1\times10^{10}$ M$_\odot$
for the STD run and
$0.4\times10^{10}$ M$_\odot$ for the DYN one.
(For all the other models see Table \ref{table:stars_1kpc}.)

To explain the origin of this difference, we separate the stars that formed
in the central galaxy of the most massive progenitor halo (second row in Figure \ref{fig:rad_distr_stars}) from
those that formed in other haloes (third row).
In the STD model, nearly 66 per cent of the stars in 
the first quintile of the age distribution ($z>4$) does not from in the most massive progenitor. On the other 
hand, in the DYN simulation, only 8 per cent of the 
final stellar mass is created at such early times
mostly in the central galaxy of the most massive progenitor (41 per cent)
and in other haloes (59 per cent).
\begin{figure}
  \includegraphics[scale=0.4]{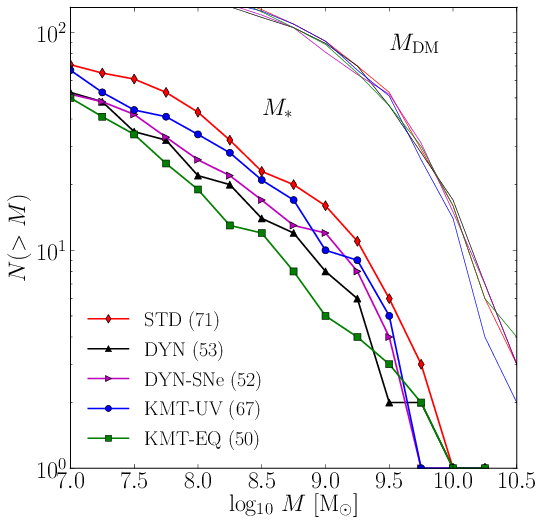}
  \caption{\label{fig:mass_function2} Cumulative distribution of the DM 
    (thin lines) and stellar mass (connected points) of substructures within the 
    virial radius of the most massive halo at $z=2$ for the STD (red), DYN (black), DYN-SNe 
    (magenta), KMT-UV (blue) and KMT-EQ (green) runs. The total number of substructures 
    (with $r<r_{\rm 200}$ and hosting at least 100 stellar particles) is indicated in parenthesis.}
\end{figure}
A visual impression can be formed from Figure \ref{fig:snapshot_z_5} where we compare the spatial distribution of DM and stars extracted from the two simulations in the region surrounding the most massive progenitor at $z=5.5$.

At this redshift, in the STD run, 99 per cent of the haloes with $M>10^9$ M$_\odot$ (i.e. above the mass threshold where SF is suppressed by the cosmic reionization of hydrogen) contain some stars while only 33 per cent do in the DYN simulation. This striking difference is due to the fact that most low-mass haloes (with $M\lsim10^{10}$ M$_\odot$) in the DYN model cannot efficiently form molecular hydrogen at early times. In fact, their dust content and central densities are too low to promote efficient H$_2$ formation, resulting in much lower SFRs. We find that their characteristic H$_2$ formation time, $\tau_{\rm f}=1/(\mathcal{R}_f\,\langle n_{\rm H}\rangle)$, is several orders of magnitude larger than the age of the Universe. Therefore, many low-mass haloes stay gas rich but dark
\citep[similar conclusions have been reached using the KMT-EQ model by][]{Kuhlen+12, Kuhlen+13,Jaacks+13,Thompson+14}.

This phenomenon has strong implications for the assembly of the galaxy in our simulations.
In the STD run, the main-progenitor galaxy is built up by a sequence of mergers between objects with extended stellar components.
On the other hand, in the DYN simulation, the same mergers provide gas-rich haloes with a few compact stellar distributions.
Once a well defined central galaxy is in place, the merger history in the two models is nearly identical. However, the
different rules with which SF and feedback take place produce a clumpier stellar pattern in the DYN model while the STD run more often develops grand design spiral arms in between major merging events.
Off-centre mergers between substructures produce large clumps in both models (see Figure 8).  In particular, a massive ($M_{\rm tot}=4\times 10^{10}$ M$_\odot$) gas-rich condensation forms at $z=2.6$ is the DYN model and orbits within a few kpc from the galactic centre afterwards. A first encounter between this object and the central stellar clump of the main galaxy takes place at $z=2.4$.  This event heats the distribution of stellar orbits in the core of the DYN galaxy and scatters the old stars (that were previously concentrated in the middle) over a more extended area (see the bottom right panel in Figure \ref{fig:rad_distr_stars}). SF in the central regions of the galaxy is also boosted at this time (see the right-hand panel in the second row of Figure \ref{fig:rad_distr_stars}).
In both models, the late history of the galaxy is characterized by a close encounter with a satellite at $2.1<z<2.2$ which mildly distorts the gas and stellar distributions.

Finally, we focus our attention on the distribution of satellite galaxies at $z=2$. Figure \ref{fig:mass_function2} shows, for all models, the cumulative number of substructures that are found within the virial radius of the main halo as a function of their DM and stellar masses.
While the number of DM clumps is virtually identical in all runs, the STD model produces a larger number of bright satellites than any other run (approximately 30 per cent more than in the DYN simulation and 10 to 30 per cent more in the other models). This is a direct consequence of the suppression of SF in metal-poor, low-mass haloes that we have already discussed above. We conclude that molecule-regulated SF can act as an effective feedback mechanism that may alleviate the satellite problem of the CDM scenario \citep[e.g.][]{Klypin+99,Moore+99}. Further study will be necessary to determine whether this conclusion also applies to low redshift galaxies.

\section{Conclusions}\label{sec:discussion}

We have presented a sub-grid model for tracking the non-equilibrium abundance of molecular hydrogen 
in cosmological simulations of galaxy formation. The novelty of the model is that it
phenomenologically 
accounts for the distribution of unresolved sub-grid densities determined from observations and simulations of the turbulent ISM. In this sense, it improves upon previous
time-dependent schemes \citep[e.g.][]{Gnedin+09,Christensen+12} in which the H$_2$ formation rate on dust grains is amplified by a fixed clumping factor.

We have implemented our model in the {\sevensize{RAMSES}} code in order to run simulations that
track the evolution of the H$_2$ content of a massive galaxy at $z=2$, and to study the imprint
of H$_2$-regulated SF.
The resulting H$_2$ fractions in different environments spanning a vast range of gas metallicities and interstellar UV fields are consistent with observations of the Milky Way Galaxy, the LMC and the SMC.

In order to better understand what determines the properties of a galaxy, we ran a suite of simulations 
of the same DM halo, each with a different prescription for computing the H$_2$ distribution.
In the runs where H$_2$ is calculated explicitly, SF was regulated by the local 
H$_2$ abundance, while, for another (STD), we adopted the traditional Schmidt law based on the total gas density.
Our main findings can be summarized as follows:

\begin{enumerate}

\item All simulations produce a galaxy which is in broad agreement with several high-redshift observational data sets. However, the different models for star and H$_2$ formation result in important differences in the stellar and gas masses of the galaxies (see Table \ref{table:par_r200}).

\item In particular, if SF is molecule-regulated and the H$_2$ abundances are computed with a detailed treatment of photo-dissociation including a simplified radiative-transfer scheme:
\begin{enumerate}
 \item The galaxy produces less stars and is in better agreement with the observed stellar-to-halo mass relation with respect to the STD model (which, however, cannot be ruled out as we did not study the effect of increasing the spatial resolution
of the simulations).
 \item The galaxy harbours a larger gas reservoir and its gas fraction better matches observations of high redshift galaxies.
 \item Early SF is inhibited in  metal-poor haloes with mass $M\lsim10^{10}$ M$_\odot$ in which gas and dust densities are too low to trigger efficient conversion of \HI~into H$_2$. 
 As a result, the main galaxy assembles by accreting many gas-rich substructures and, consequently, hosts a younger stellar population and harbours a larger cold gas
reservoir than in the STD case.
 Also, the number of satellites of the main galaxy at $z=2$ is reduced by 30 per cent at all stellar masses compared with the STD simulation.
\end{enumerate}

\item Regardless of the H$_2$ model:
\begin{enumerate}
 \item The main galaxy in our simulations (with the exception of the KMT-EQ model) has similar spatial distribution and total mass of molecular hydrogen at $z=2$. This is mainly due to the fact that the average metallicity of the gas in its dominant progenitors is already $0.1$ Z$_\odot$ at $z=9$. As a result, H$_2$ formation is rapid, mitigating any subtle differences independent of whether stars form from atomic or molecular gas.
 \item The molecular mass in a cell scales linearly with that of the metals 
  (above a model-dependent threshold density). This is a consequence of assuming that dust traces the metals 
  in the simulations.
\end{enumerate}

\item If H$_2$ destruction by SNe is substantial, SF is suppressed by nearly 20 per cent at all times relative to an identical model in which this destruction channel is neglected.

\item Contrary to the assumption that gas is fully molecular in high-redshift galaxies (commonly used to 
  interpret CO observations, e.g. \citet{Genzel+10,Tacconi+10,Magnelli+13}), the atomic gas fraction in 
  our simulated galaxy is comparable to the molecular contribution, independent of the H$_2$
  formation model.
  
  \item Using the STD model to form stars, but `painting on' H$_2$ in post-processing using the KMT-UV prescription, gives a reasonable estimate of the total H$_2$ mass of the galaxy. However, as already mentioned above, the resulting galaxy contains more stars (in particular old stars) than in all of the molecule-regulated schemes.

\end{enumerate}

Although our non-equilibrium H$_2$ model represents a significant improvement over previous 
methods, many challenges remain. For instance, we have assumed that the sub-grid density PDF 
of GMCs can be accurately described by a lognormal distribution. This is based on several observations of 
molecular clouds which, in some cases, show high-density tails in star forming regions 
\citep[e.g.][]{Kainulainen+09,Schneider+13}. Several complex physical phenomena like energy injection,
turbulence, gravity and external compression influence the density structure of molecular clouds. 
Yet, numerical studies of the ISM have shown that the lognormal model is a good approximation when 
an isothermal gas flow is supersonically turbulent 
\citep[e.g.][]{Vazquez-Semadeni+94,Glover+07a,Glover+07b,Federrath+09,Federrath+13}.
Power-law tails in the high-density regime form under the presence of  self-gravity (which generates 
dense cores and super-critical filaments). Non-isothermal turbulence can also increase the occurrence 
of dense clumps. None the less, these uncertainties are likely sub-dominant to those associated with 
modelling the effects of SNa feedback on GMCs, which must be tackled with small-scale simulations of
the ISM.

In addition, we have set the dispersion of the lognormal density distribution 
to be $\sigma\simeq 1.5$, consistent with a constant clumping factor $C_\rho=10$. This choice was
motivated by theoretical work that relates local density enhancements to the three-dimensional 
rms Mach number, $\mathcal{M}$, with values of $\mathcal{M}\sim 5.5$ \citep[e.g.][]{Padoan+97,Ostriker+01,Price+11}. 
Moreover, the same value for the clumping factor has been adopted in the literature to best match 
the H$_2$ content in observations and simulations \citep[e.g.][]{Gnedin+09,Christensen+12}.
However, observations of GMCs have revealed substantial variations in the Mach number 
\citep{Schneider+13}. In future implementations, the realism of our model can be improved by 
adjusting the clumping factor (as well as the PDF of the sub-grid  density) in cells with different 
mean densities and temperatures. From the technical point of view, this is straightforward to do:
the difficulty lies in linking the mean properties of a cell to the sub-grid parameters that regulate 
the density PDF. One intriguing possibility could be to implement a simplified description of supersonic 
turbulence along the lines of that proposed by \citet{Teyssier+13}. 
We plan to return to these issues
in future work.

\section*{Acknowledgements}
We wish to thank our referee for a constructive report that has
improved this paper. We also thank Romain Teyssier and Tom Abel for helpful discussions and acknowledge support
from Mark Labadens for the software PyMSES and from Oliver Hahn for the MUSIC code.
We are also grateful to Andreas Schruba for kindly providing us with data from the HERACLES survey.
This work was supported by the Deutsche Forschungsgemeinschaft (DFG) through the 
project SFB 956 {\it Conditions and Impact of Star Formation}, sub-project C4.
MT was supported through a stipend from the International Max Planck Research School 
(IMPRS) for Astronomy and Astrophysics in Bonn. We acknowledge that the results of this 
research have been achieved using the PRACE-2IP project (FP7 RI-283493) resources HeCTOR
based in the UK at the UK National Supercomputing Service and the Abel Computing Cluster 
based in Norway at the University of Oslo.

\bibliographystyle{mn2e}

\appendix

\section{Interstellar UV field}\label{sec:uv}

Each star particle in our simulations represents a simple stellar
population. We use the STARBURST99
templates \citep{Leitherer+99} -- with a \citet{Kroupa+01} initial mass
function (IMF) --
to compute the luminosity in the LW band, $L_{\rm
  LW}(t_s)$, as a function of stellar age, $t_s$.
We then
propagate this radiation up to a maximum distance assuming that the gas
is optically thin. For each spatial location,
we only consider sources of radiation that lie within an
oct, i.e. a collection of 8 cells at the highest level of refinement.
Therefore the corresponding UV flux is:

\begin{equation}
G(t) = \frac{1}{G_0^\prime}\,
\alpha\frac{\sum_{i\in{\rm oct}}L_{\rm LW}(t-t_{s,i})}{\,4\pi\,\Delta x^2\,\Delta\lambda\,c}\,\,\,\,,
\end{equation}
where $t$ is the time elapsed in the simulation, $\Delta x$ is the size of a resolution element, $c$ is the speed of light, $\Delta\lambda=1110-912$ \AA~and $G_0^\prime=4\times10^{-17}$ erg cm$^{-3}$ \AA$^{-1}$.
The factor $\alpha\simeq 2.876$ gives the correct average flux at the
centre of a cell when stars are randomly distributed (with uniform density)
within an oct. This coefficient has been measured with a Monte Carlo method.
In our simulations, $\Delta x\simeq 180$ pc and is comparable with the 
typical size of GMCs.

\label{lastpage}

\end{document}